\documentstyle[psfig,emulateapj]{article}
\def\etal   {{et~al.}\ }
\def\kms{{\rm\,km/s}}
\def\msun{{\rm\,M_\odot}}

\def\vol#1  {{{#1}{\rm,}\ }}
\def\lya{{\rm Ly}\alpha}

\def\etal{et al.\ }

\def\clock{\count0=\time \divide\count0 by 60
     \count1=\count0 \multiply\count1 by -60 \advance\count1 by \time
     \number\count0:\ifnum\count1<10{0\number\count1}\else\number\count1\fi}

\font\eightrm=cmr8 scaled \magstep0
\def\fig #1, #2, #3, #4, #5, #6 {
\topinsert
\smallskip
\centerline{\psfig{figure=#1,height=#2 in,width=#3 in,angle=#4}}
\medskip
{\vskip #5 cm\leftskip2.5em \parindent=0pt {\eightrm #6 }}
\endinsert}

\begin{document}
\title{Testing Cold Dark Matter Models At Moderate to High Redshift}
\author{Renyue Cen\altaffilmark{1}}
\altaffiltext{1} {Princeton University Observatory, Princeton University, Princeton, NJ 08544; cen@astro.princeton.edu}

\begin{abstract}

The COBE microwave background temperature fluctuations 
and the abundance of local rich clusters of galaxies
provide the two most powerful constraints on cosmological models.
When all variants of the standard cold dark matter (CDM) model
are subject to the combined constraint,
the power spectrum of any model is fixed to $\sim 10\%$ accuracy
in both the shape and overall amplitude.
These constrained models are not expected to differ dramatically 
in their local large-scale structure properties.
However, their evolutionary histories differ,
resulting in dramatic differences towards high redshift.
In particular, it should be true that any statistical measure
that probes a rapidly diminishing tail of some distribution should 
provide a sensitive test at some sufficiently high 
redshift, when the objects in question are rare and hence in the tail.

We examine in detail six standardized,
COBE and cluster normalized CDM 
models with respect to a large set of independent observations.
The observations include
correlation function of rich clusters of galaxies,
galaxy power spectrum,
evolution of rich cluster abundance,
gravitational lensing by moderate-to-high
redshift clusters,
$\lya$ forest,
damped $\lya$ systems,
high redshift galaxies,
reionization of the universe and
future CMB experiments.
It seems that each of the independent observations
examined is or potentially is capable of  
distinguishing between at least some of the models.
The combined power of several or all of these observations 
is tremendous.
Thus,
we appear to be on the verge 
of being able to make dramatic tests of all models
in the near future using a rapidly growing set of 
observations, mostly at moderate to high redshift.
Consistency or inconsistency between different observed
phenomena on different scales and/or at different epochs
with respect to the models
will have profound implications 
for theory of growth of cosmic structure.
\end{abstract}


%\notetoeditor{}

\section{Introduction}

On one hand,
the success of the COsmic microwave Background Explorer 
(COBE; Smoot \etal 1992) can hardly be overstated;
it revolutionized our way of quantifying cosmology.
On the other hand, it did little justice to enthusiastic cosmologists:
rather than helping us nail down to an ultimate model,
it proded us to introduce more models (or variants)
to meet with the new observational constraint.
Post-COBE era is filled with new ideas about models
at the expense of introducing more parameters to the 
simple, non-flavored, standard cold
dark matter model (Peebles 1982; Blumenthal \etal 1984), 
which has been remarkably successful
in matching a large set of local observations for the past decade 
(e.g., Davis  \etal 1985; White \etal 1986; Frenk \etal 1988; 
Park 1990; Cen \& Ostriker 1992a and references therein),
although some evidence of its
lack of large-scale power began to emerge 
(e.g., Maddox \etal 1990) in the early nineties
(see Ostriker 1993 for an excellent review).
The flourish of new models is largely thanks to 
the general consensual requirement that 
any cosmological model 
match both the COBE observation on large scales at an early epoch
and local observations, primarily of distributions of galaxies. 

The newly introduced models that are deemed to be interesting
are mostly based on the general success of the standard
cold dark matter model.
The variants arise by introducing 
one or two more parameters
to the standard critical density CDM model with Zel'dovich
primordial spectrum: the shape of the power spectrum (characterized
by the primordial power index on the large scales, $n$)
and/or the composition of the universal matter
including vacuum energy density (cosmological constant).
It should be noted that 
variations of the latter also change  the overall
shape of the power spectrum, quite separate from the primordial spectrum.
The consequence of allowing such added freedom is
the current state of a degenerate set of 
``viable" COBE and cluster-normalized CDM models typified by the 
tilted cold dark matter model 
(\cite{cgko92}, \cite{lls92}; \cite{lc92}; \cite{a93}; \cite{lmm93}),
the mixed cold dark matter model 
(\cite{dss92}; \cite{tr92}; \cite{k93}; \cite{co94}; \cite{mb94}),
the cold dark matter model with a non-zero cosmological constant
(\cite{p84};
\cite{ebw92};
\cite{kgb93};
\cite{cgo93};
\cite{os95})
and the open cold dark matter model (Gott 1982; Bucher, Goldhaber, \& Turok 1995).
Although noticeably being paid significantly less attention to,
the primeval baryon isocurvature model (PBI,
Peebles 1987a,b; Cen, Ostriker, \& Peebles 1993)
still appears contentious, especially with its 
virtue of early structure formation.

After a general world model is chosen (i.e., $\Omega_0$, $\Lambda_0$, $H_0$
and dark matter and baryonic compositions are chosen),
the shape and the overall amplitude 
of its power spectrum are yet to be
fixed before the model becomes completely deterministic.
COBE provides one fixing point at large scales ($\lambda\sim 1000h^{-1}$Mpc).
The abundance of local rich clusters of galaxies 
provides another fixing point at intermediate scales
($\lambda\sim 10h^{-1}$Mpc).
Both points are currently determined to an accuracy of $\sim 10\%$.
Once the power spectrum is also fixed, the entire 
history of a model universe can, in principle, be computed.
Sufficiently accurate numerical techniques 
with sufficiently realistic physical modeling 
have become more available due to parallel advances
in both computer hardware and software capabilities
to allow for accurate predictions for at least some 
aspects or phenomena of the cosmic structure formation
to be made, which can be compared to 
observations on various scales and at various epochs.
Examples are $\lya$ clouds (Cen \etal 1994; Zhang \etal 1995;
	Hernquist \etal 1996; MCOR),
X-ray clusters (Kang \etal 1994; Bryan \etal 1994; Cen \& Ostriker 1994a;
Frenk \etal 1998)
and gravitational lensing by clusters of galaxies
and large-scale structure 
(Cen \etal 1994; Wambsganss \etal 1995).

Parallel in time,
rapid advances in observational capabilities in recent
years, thanks in large part to the ground-based 10-meter class
Keck telescopes as well as the Hubble Space Telescope (HST),
make possible explore regimes which were previously unaccessible.
For example, HST opens a brand new window to studying
the formation and evolution of galaxies to a much higher
redshift than before (e.g., Lowenthal \etal 1997)
and quasar absorption systems
at some entirely new wavelength bands (e.g., Bahcall \etal 1996).
Observations by Keck 
telescopes have made possible 
routinely record high quality quasar absorption
spectra (e.g., Hu \etal 1995), providing new opportunities 
to conduct detailed studies of Lyman alpha forest lines and
related metal line systems (e.g., Lu \etal 1997).
Observations of Lyman break galaxies using Keck 
telescopes have opened yet another new window to 
the universe at high redshift (Steidel \etal 1998, S98 henceforth).
Deeper and larger surveys of X-ray clusters of galaxies 
as well as optically identified clusters of galaxies
are pushing the redshift of observed rich,
luminous and hot clusters
to beyond unity (e.g., Rosati \etal 1998).
More excitingly, more and rapid observational advances 
are expected to take place
in the near future in anticipation of  
many large projects and next generation telescopes
convering a variety of wavelengths from meterwave
to $\gamma$-ray.

This paper attempts to interface between these two developments.
To set a standard for comparing different models 
and for comparing results between different workers in the field,
we adopt the set of six CDM models 
presented in Cen (1998; C98 henceforth) 
for detailed analyses.
All the models are normalized to both COBE on
large scales and the observed abundance of local rich clusters 
of galaxies,
thus titled ``viable".
The goal of this paper
is to address the opportunities that new observations are and will soon be
providing us with new ways to test
these remaining, ``viable" models.
It is hoped that some theoretical insight may be extracted
to help highlight some potentially promising observations
for (at least) the purpose of testing cosmological models.
It is exciting that each of the independent observations
examined here, mostly at moderate-to-high redshift,
seems to be or potentially be capable of  
distinguishing between at least some of the models.
The combined power of several of these observations 
is tremendous, enabling the possibility of 
nailing down the ultimate cosmological model with 
high precision.  
It is perhaps more important 
that different observations at completely
different scales and/or epochs
should provide an overall consistency check of 
the general gravitational instability picture 
and/or Gaussianity of the primordial density fluctuations.

The main results of the paper are organized as follows.
The parameters of a suite of six 
CDM models that are normalized to COBE
and local cluster abundance
are discussed in \S 2. 
\S 3 presents tests on the models
with nine independent observations 
in order of ascending redshift in the subsections:
the cluster correlation function (\S 3.1),
the galaxy power spectrum from
the Sloan Digital Sky Survey (SDSS) (\S 3.2),
the evolution of rich cluster abundance (\S 3.3),
the gravitational lensing by moderate-to-high
    redshift clusters of galaxies (\S 3.4),
the $\lya$ forest (\S 3.5),
the damped $\lya$ systems (\S 3.6),
the high redshift galaxies (\S 3.7),
the reionization of the universe (\S 3.8) and
the future CMB experiments (\S 3.9).
Discussion of the constraining/differentiating power and observational
requirement of each of the suite of observations 
is presented in \S 4.
Conclusions are given in \S 5.

\section{Six COBE and Cluster Normalized CDM Models}

\begin{deluxetable}{cccccccccc} %{l,r}
\tablewidth{0pt}
\tablenum{1}
\tablecolumns{10}
\tablecaption{Six COBE and Cluster-Normalized CDM Models} %\label{tab1}}
\tablehead{
\colhead{Model} &
\colhead{$H_0$} &
\colhead{$n$} &
\colhead{$\Omega_c$} &
\colhead{$\Omega_h$} &
\colhead{$\Lambda_0$} &
\colhead{$\Omega_b$} &
\colhead{$\sigma_8$}}

\startdata
tCDM & $55$ & $0.77$ & $0.936$ & $0.00$ & $0.0$ & $0.064$ & $0.55$ \nl 
HCDM & $55$ & $0.88$ & $0.736$ & $0.20$ & $0.0$ & $0.064$ & $0.52$ \nl 
OCDM25 & $65$ & $1.47$ & $0.220$ & $0.00$ & $0.0$ & $0.030$ & $1.00$ \nl 
OCDM40 & $60$ & $1.15$ & $0.346$ & $0.00$ & $0.0$ & $0.054$ & $0.80$ \nl 
$\Lambda$CDM25 & $65$ & $1.10$ & $0.220$ & $0.00$ & $0.75$ & $0.030$ & $0.95$ \nl 
$\Lambda$CDM40 & $60$ & $0.96$ & $0.346$ & $0.00$ & $0.60$ & $0.054$ & $0.80$ \nl 
\enddata
\end{deluxetable}

\begin{figure*}
\centering
\begin{picture}(400,200)
\psfig{figure=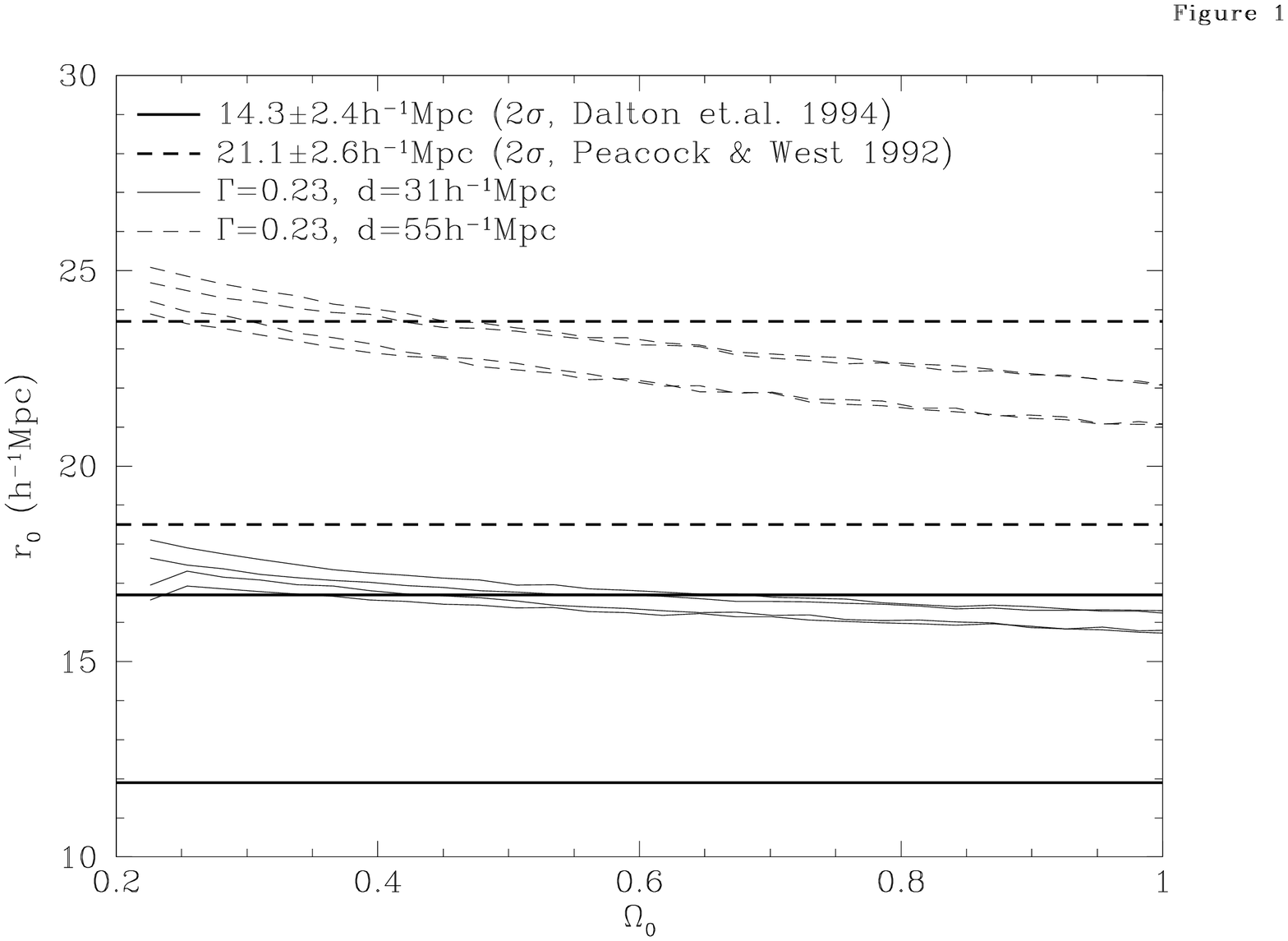,height=10.0cm,width=15.0cm,angle=270.0}
\end{picture}
\caption{
Cluster-cluster two-point correlation length
computed using GPM
as a function of $\Omega_0$ for clusters
with mean separation of $55h^{-1}$Mpc (light dashed curves)
and $31h^{-1}$Mpc (light solid curves), respectively.
For all models we use
the CDM transfer function from BBKS
with $\Gamma\equiv \Omega_0 h=0.23$ and $n=1.0$.
All models are normalized to clusters using equation 36 of C98.
Each set of four curves consists of two upper curves
using $2\sigma$ upper limit of $\sigma_8$ by equation 36 of C98
with $\Lambda_0=0$ and with $\Omega_0+\Lambda_0=1$,
and two lower curves using $2\sigma$ upper limit of $\sigma_8$ 
with $\Lambda_0=0$ and with $\Omega_0+\Lambda_0=1$.
Also shown as two heavy dashed horizontal line
indicate the $2\sigma$ upper and lower bounds from observations
by Peacock \& West (1992)
of the
correlation length of clusters of mean separation of $55h^{-1}$Mpc.
The two heavy solid horizontal lines
are the $2\sigma$ upper and lower bounds from observations
by Dalton \etal (1994) of the
correlation length of clusters of mean separation of $31h^{-1}$Mpc.
}
\end{figure*}

The models are taken from C98 (Table 4 of that paper)
and listed in Table 1.
When normalized to both COBE on very large scales
and the abundance of local rich clusters of galaxies,
both the shape ($n$) and amplitude of 
the power spectrum ($\sigma_8$) of any model 
are fixed to about 10\% accuracy (C98).
These six models are determined to this accuracy.
The tCDM model has $\Omega_0=1$ but with a tilt on the primordial
power spectrum.
The HCDM model has $\Omega_0=1$ with 20\% mass
in massive neutrinos.
Two neutrino species of equal mass are assumed.
OCDM25 and OCDM40 are two open models
and 
$\Lambda$CDM25 and $\Lambda$CDM40 are
two low density but flat models with a 
cosmological constant.
Except the open models, a tensor component is added when
normalizing each model to COBE with
tensor to scalar ratio $T/S=7(1-n)$ 
(Liddle \& Lyth 1992; Davis \etal 1992;
Crittenden \etal 1993;
Stewart \& Lyth 1993).
Note that some tilt is present in all models listed,
required to match both COBE and local cluster abundance.
This set of six CDM models 
likely brackets all potentially interesting models.

The baryon densities for OCDM25 and $\Lambda$CDM25 
are computed using $\Omega_b h^2=0.0125$ (Walker \etal 1991),
and for the remaining four models
using $\Omega_b h^2=0.0193$ (Burles \& Tytler 1997).
These choices of $\Omega_b$ for the models
are made to maximize the viability of each model
with respect to the observed gas fraction
in X-ray clusters of galaxies (White \etal 1993b;
Lubin \etal 1996;
Danos \& Pen 1998, $\rho_{gas}/\rho_{tot}=(0.053\pm 0.004)h^{-3/2}$).
The power spectrum transfer functions for all the models
are computed using CMBFAST code developed by Seljak and Zaldarriaga.
The choice of the Hubble constant is made for each model such that
each model is consistent with current measurements of
the Hubble constant.
It appears that $H_0(obs)=65\pm 10$km/s/Mpc can account for
the distribution of the current data from various measurements
(see, e.g., Trimble 1997),
except for those from Sunyaev-Zel'dovich observations
(for a discussion of 
a reconciliation of this difference, see C98).
Another consideration is that
the age constraint
from latest globular cluster observations/interpretations
(c.f., Salaris, Degl'Innocenti, \& Weiss 1997)
is not violated.

\section{Testing CDM Models At Moderate to High Redshift}

While COBE and local cluster abundance combine to provide
an unprecedented accuracy of fixing models in terms of
limiting the parameter space,
some degeneracy remains.
It is therefore
essential to examine other, independent
observations to help further tighten the parameter space.
We will examine various observations 
in this section using the six models listed in Table 1 
as a testbed.
We consider nine independent observations 
in order of ascending redshift:
the cluster correlation function at low redshift (\S 3.1),
the galaxy power spectrum from SDSS (\S 3.2),
and evolution of rich cluster abundance (\S 3.3),
the gravitational lensing by moderate-to-high
    redshift clusters of galaxies (\S 3.4),
the $\lya$ forest (\S 3.5),
the damped $\lya$ systems (\S 3.6),
the high redshift galaxies (\S 3.7),
the reionization of the universe (\S 3.8) and
the future CMB experiments (\S 3.9).

\subsection{Constraints by the Cluster Correlation Function at $z=0$}

\begin{figure*}
\centering
\begin{picture}(400,250)
\psfig{figure=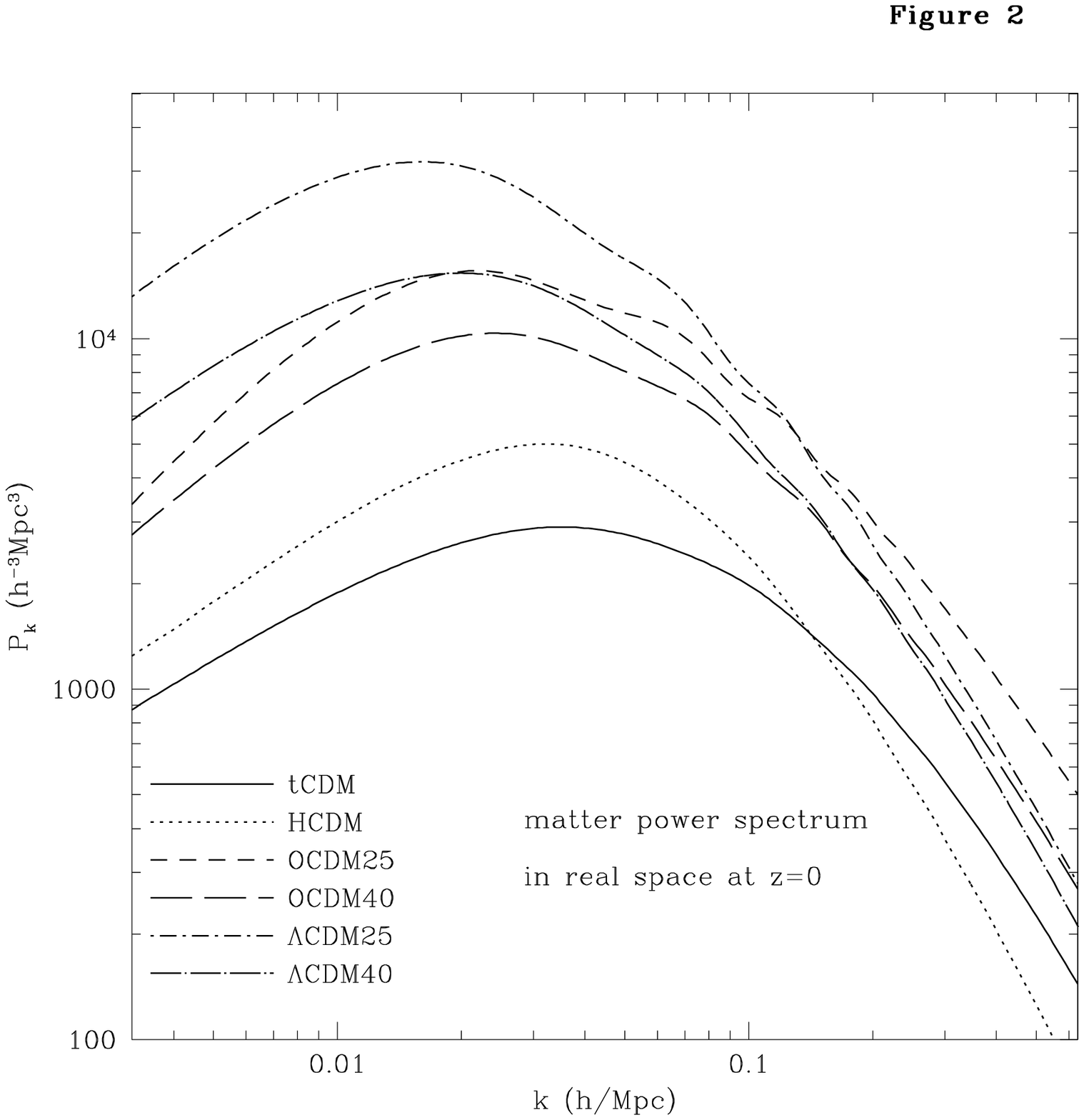,height=10.0cm,width=15.0cm,angle=0.0}
\end{picture}
\caption{
shows the power spectra (in real space) for the six models in Table 1.
}
\end{figure*}
\begin{figure*}
\centering

\begin{picture}(400,250)
\psfig{figure=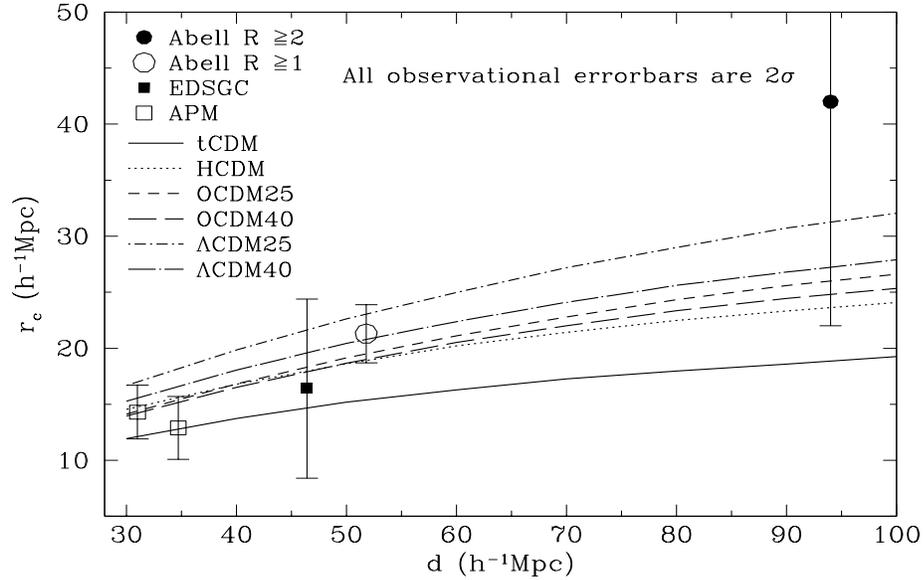,height=10.0cm,width=15.0cm,angle=270.0}
\end{picture}
\caption{
shows the correlation length
against the mean separation of the clusters in the six models (Table 1).
Shown as symbols are observations from various sources,
where
the filled dot is for $R\ge 2$ Abell clusters
from Bahcall \& Soneira (1983) with separation $d=94h^{-1}$Mpc
and correlation length $r_0=42\pm 20h^{-1}$Mpc,
the open circle is for $R\ge 1$ Abell clusters
from Peacock \& West (1992) with separation $d=51.8h^{-1}$Mpc
and correlation length $r_0=21.3\pm 2.6h^{-1}$Mpc,
the solid square is for Edinburgh-Durham Southern Galaxy Catalogue (EDSGC)
from Nichol \etal (1992) with separation $d=46.4h^{-1}$Mpc
and correlation length $r_0=16.4\pm 8.0h^{-1}$Mpc,
and
the two open squares are for APM Galaxy survey
from Dalton \etal (1992) with separation $d=34.7h^{-1}$Mpc
and correlation length $r_0=12.9\pm 2.8h^{-1}$Mpc
and
from Dalton \etal (1994) with separation $d=31.0h^{-1}$Mpc
and correlation length $r_0=14.3\pm 2.4h^{-1}$Mpc.
All the observational errorbars are $2\sigma$.
}
\end{figure*}

Before examining the correlation functions of the six models,
it is worthwhile to understand factors
that determine the cluster correlation function.
Figure 1 shows the cluster-cluster
two-point correlation length (i.e., the scale at which
the correlation function is unity), computed using 
the Gaussian Peak Method (GPM, C98),
as a function of $\Omega_0$ for clusters
with mean separation of $55h^{-1}$Mpc (light dashed curves)
and $31h^{-1}$Mpc (light solid curves), respectively.
For all models shown here, we use $\Gamma=0.23$
with the CDM transfer function from BBKS and
normalize to the local cluster abundance (equation 36 of C98)
but not necessarily to COBE.
From Figure 1 it immediately becomes clear 
that the cluster correlation function almost does not depend on
$\Lambda_0$ at all, only weakly depends on $\Omega_0$ and 
somewhat moderately depends on $\sigma_8$:
$dr_0/d\Lambda_0\approx 0$,
$dr_0/d\Omega_0\approx 2-4$,
and
$dr_0/d\sigma_8\approx 6-10$.
Therefore, given the uncertainties in fixing $\sigma_8$ of a model,
uncertainties in the shape of the power spectrum
and uncertainties in the observed correlation length of
clusters, it is not possible
to constrain $\Omega_0$ using correlation functions of clusters
of galaxies, 
{\it unless models with different $\Omega_0$ 
also substantially differ in $P_k$ on the relevant large scales 
($30-300h^{-1}$Mpc)}.

Fortunately, the power spectra of various models
with different $\Omega_0$/$\Lambda_0$ are indeed quite different
(Table 1),
as required to fit both COBE and local rich cluster abundance.
Figure 2 shows the power spectra (in real space)
for the six models in Table 1.
It is seen that normalizing to both COBE and clusters
results in a wide disparity in shape and amplitude of the power
spectra of the models in a wide range of scales. 
The correlation length of clusters of galaxies in the six models
are then shown in Figure 3 in a somewhat different manner,
where the correlation length in each model is plotted
against the mean separation of the clusters in question.
Note that the mean separation of 
a set of clusters is
the mean separation of a set of 
the richest (most massive) clusters in a model;
clusters are rank-ordered in mass and selected in descending order in mass.
We see that there exists significant differences
between different models shown here.
Since we have 
explicitly shown earlier
that the dependence of the correlation length on $\Omega_0$ and $\Lambda_0$
is rather weak, the differences in the correlation lengths
between the models shown here chiefly attribute to the differences
in the power on large scales between the models.
The trend is clear and physically understandable:
the more power a model has on large scales $\lambda \geq 100h^{-1}$Mpc,
the stronger dependence of the correlation length on the
richness of the clusters (see Figure 2 to compare
the power spectra of the models) is displayed,
consistent with previous findings (Bahcall \& Cen 1992;
Croft \& Efstathiou 1994).

It is evident that the correlation length of clusters of galaxies
sets a useful constraint on models.
If the measurement accuracy
can reach $\le 1h^{-1}$Mpc,
one can potentially strongly differentiate between models.
It is also clear
that it will be more useful to reduce the observational errorbars 
of the correlation length of very rich clusters (e.g., 
$d\ge 90h^{-1}$Mpc)
for the purpose of testing models,
since the models differ most there.
Stated somewhat more quantitatively in a hypothetic sense,
were the $2\sigma$ errorbar of the solid dot 
in Figure 3 smaller than $5h^{-1}$Mpc, 
all the models considered here are ruled out at
a very high confidence level ($\sim 4\sigma$).

Consistent with other observations (e.g., Maddox \etal 1990)
and previous results of clustering analyses of galaxy clusters 
(e.g., Bahcall \& Cen 1992; Croft \& Efstathiou 1994),
the standard $\Omega_0=1$ CDM model 
even with a significant tilt of the power spectrum ($n=0.77$, tCDM)
is in disaccord with 
currently available observations due to the lack of large-scale power.
HCDM model and all other, low density
models are consistent with available observations of
cluster correlations.

\subsection{Galaxy Power Spectrum}

\begin{figure*}
\centering
\begin{picture}(400,300)
\psfig{figure=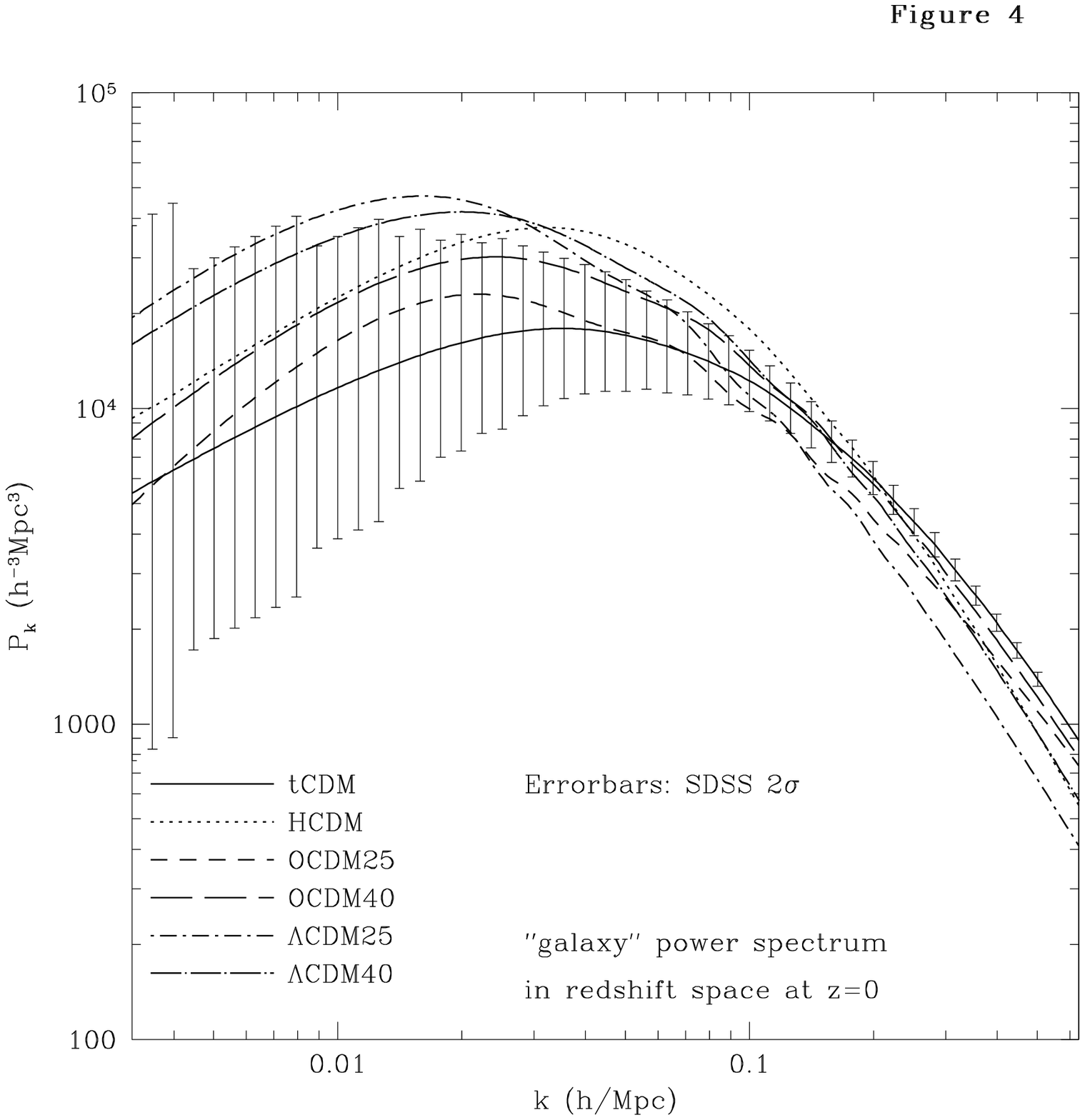,height=10.0cm,width=15.0cm,angle=0.0}
\end{picture}
\caption{
The power spectra of galaxies
{\it in redshift space} for the six models (Table 1) at $z=0$.
See text for how they are obtained.
For tCDM, we also show the errorbars ($2\sigma$)
that SDSS are thought to
achieve, adapted from Goldberg \& Strauss (1998).
}
\end{figure*}

The Sloan Digital Sky Survey now underway
(Knapp, Lupton, \& Strauss 1996) will provide 
a tremendously large database for a variety of objects.
Realizing its full value will perhaps require the work
by generations of astronomers.
One obviously valuable piece of information that can be 
extracted immediately upon the availability of 
the distribution of galaxies from SDSS,
up to very large scales
previously unreached in full three dimension.
Figure 4 shows the power spectra
of galaxies for the six models (Table 1)
{\it in redshift space} at $z=0$.
The shown galaxy power spectrum in redshift space for each model 
is related to the linear matter power spectrum in real space
by the following relation:
\begin{equation}
P_{gal,red}(k)=(1+{2\over 3}\Omega_0^{0.6}+{1\over 5}\Omega_0^{1.2})\sigma_8^{-2}P_{mass,real}(k).
\end{equation}
\noindent 
In the above relation the first term on
the right hand side  
is due to linear peculiar velocity distortion (Kaiser 1987)
and the second term on 
the right hand side  
reflects the (linear) bias of galaxies over mass in order
for each model to be consistent with the observed galaxy
density fluctuations at z=0.
For tCDM, we also show the errorbars ($2\sigma$)
that SDSS can achieve, adopted from 
Goldberg \& Strauss (1998).
Errorbars for other models are the same but not shown for readability
of the figure.
It is clear that the SDSS galaxy power spectrum
in a large range in $k$ from $0.01~h/$Mpc to $0.5~h/$Mpc
will provide some discriminatory tests of the models.
Taken at face value, it seems that
the errorbars are smallest at the highest $k$,
where the data points may provide the most discriminating 
test of the models.
But we caution that 
nonlinear effects 
may wash out some of the differences between
the models at the high $k$ end (e.g., Gramann, Cen \& Bahcall 1993).
We should only conservatively
estimate the discriminating power of 
the galaxy power spectrum from SDSS at
$k<0.2~h$/Mpc, where the nonlinear effects are relatively mild.
A close examination of Figure 4
reveals that there is not a pair of models
that have the same power at $k<0.2~h/$Mpc
within the $2\sigma$ errorbars provided by SDSS.
Therefore, SDSS should be able to test 
all these models at a confidence level $>2\sigma$.
The purpose of this investigation is to illustrate
the potential of SDSS galaxy power spectrum
in placing strong constraints on models.
Quantitative comparisons will have to be 
made between large-scale, fully nonlinear cosmological galaxy formation
simulations (Cen \& Ostriker 1992b,1993b,1998)
and observations.

\subsection{Evolution of Cluster Mass Function}

\begin{figure*}
\centering
\begin{picture}(400,300)
\psfig{figure=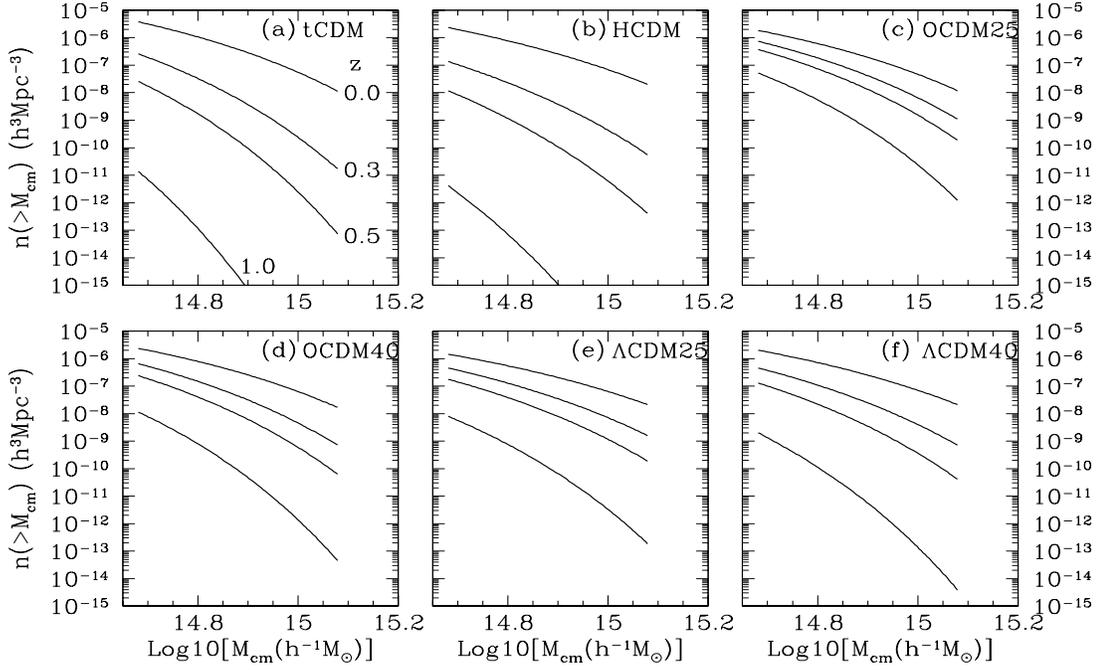,height=10.0cm,width=15.0cm,angle=270.0}
\end{picture}
\caption{
It displays the mass functions for the models
at four redshifts, $z=0.0, 0.3, 0.5, 1.0$.
The cluster mass is defined as
the mass in a sphere of {\it comoving}
Abell radius ($r_A=1.5\;h^{-1}$Mpc) at all redshifts.
All models are normalized to the observed
mass function at redshift zero.
}
\end{figure*}

Let us now study the evolution of the cluster mass functions
in the six models (Table 1),
computed using GPM (C98).
Figure 5 displays
the mass functions for the models
at four redshifts, $z=0.0, 0.3, 0.5, 1.0$.
The cluster mass is defined as 
the mass in a sphere of {\it comoving}
Abell radius ($r_A=1.5\;h^{-1}$Mpc) at all redshifts.
Again, {\it all models are normalized to the observed
mass function at redshift zero as well as COBE}.
As one would have naively expected, 
the evolution is strongest in the $\Omega_0=1$ models (5a,b)
and weakest in the open, lowest density model (5c).
It must be stressed, however,
as we will show shortly, 
that the physical reason for this trend may come as a surprise.
{\it The underlying cause is $\sigma_8$, not $\Omega_0$.}
The HCDM (5b) shows a comparable rate of evolution to tCDM
(5a) because the reduction of power on smaller scales
$k>0.12$h/Mpc is roughly compensated by
a comparable increase of power on larger scales 
$k<0.12$h/Mpc in HCDM compared to tCDM.
Other models show intermediate rates of evolution,
in a comprehensible order.
Figure 6 is similar to Figure 5 with
the only difference being that the cluster mass is now defined as 
the mass in a sphere of {\it proper}
Abell radius ($r_A=1.5\;h^{-1}$Mpc) at all redshifts.
The trend between models 
is similar in Figure 6 to that in Figure 5.
But the evolutionary rates in all models are
dramatically smaller in Figure 6 than in Figure 5.
This says that the mass of clusters within a physical
radius evolves much more slowly 
than the mass of clusters within a comoving radius.
This is trivial, since a fixed physical
radius corresponds to a larger comoving radius 
at a higher redshift thus its sphere contains more mass 
than its counterpart with the comoving radius.
A more subtle difference between the models 
is that the evolution of the mass function for the low density open model
(OCDM25)
to redshift unity (the bottom curve in Figure 6c)
is much smaller than that of any other model.
In other words,
the mass of clusters within a physical
radius in OCDM25 evolves only very slowly.
An equivalent, conventional but perhaps conceptually simpler
way to put this is that the clusters in OCDM25
virialize early at $z\sim 1$ and later mass accretion onto
the formed clusters through the surface 
of the indicated radius is modest.

\begin{figure*}
\centering
\begin{picture}(400,300)
\psfig{figure=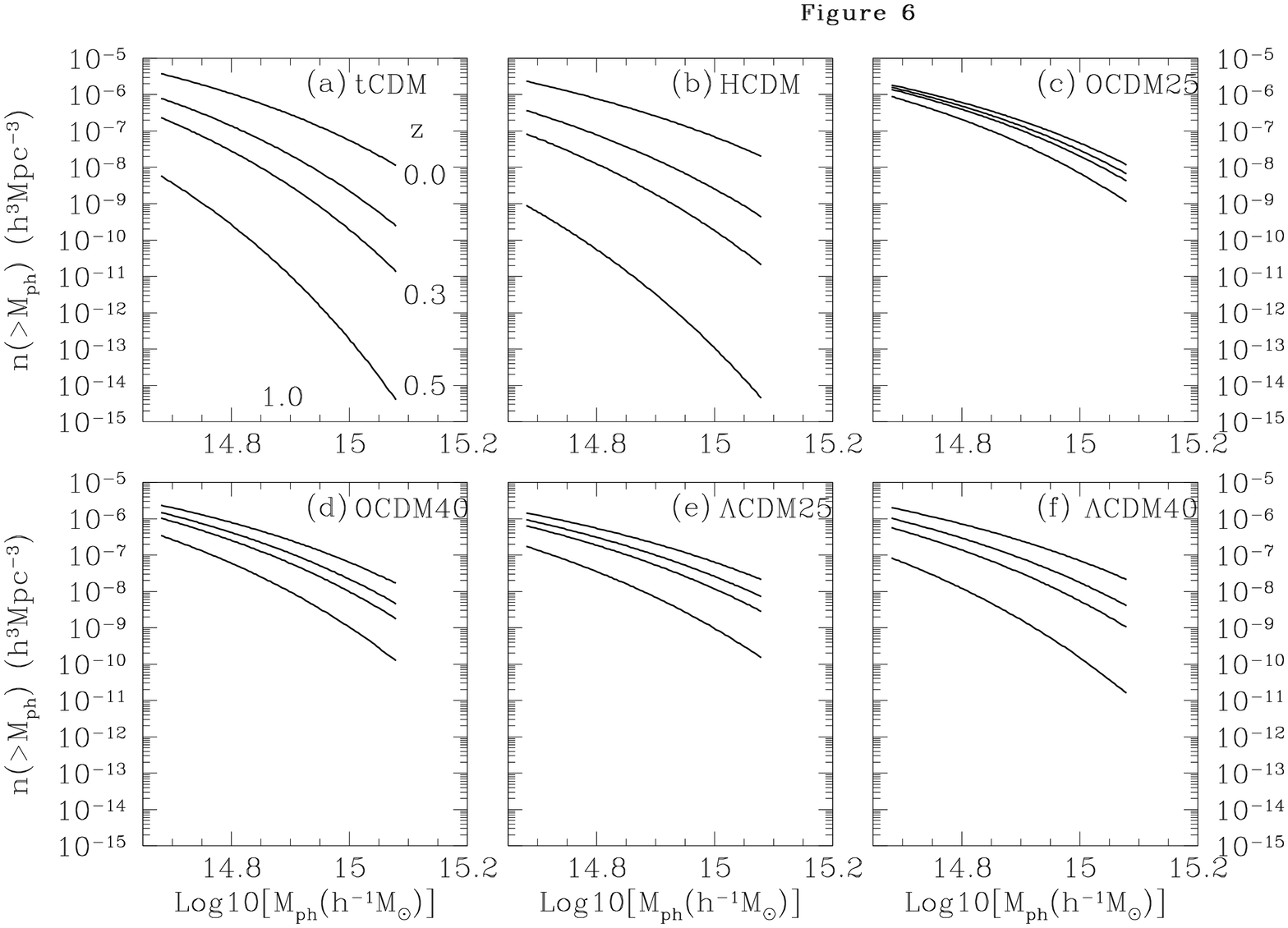,height=10.0cm,width=15.0cm,angle=270.0}
\end{picture}
\caption{
is similar to Figure 5 with
the only difference being that the cluster mass is defined as
the mass in a sphere of {\it proper}
Abell radius ($r_A=1.5\;h^{-1}$Mpc) at all redshifts.
}
\end{figure*}

We would summarize the trends for clarity.
{\it Cluster mass function is a monotonic function in time:
the abundance of clusters at a fixed mass 
(at least in the range of mass considered 
here from $4.8\times 10^{14}h^{-1}\msun$ to $1.2\times 10^{15}h^{-1}\msun$)
decreases with increasing redshift.
For all COBE and cluster (at $z=0$) normalized models,
the lower is $\Omega_0$, the weaker is the evolution;
for two models with a same $\Omega_0$, an addition of a cosmological constant
enhances the evolution,}
especially beyond a certain high
redshift when the values of $\Omega_z$ in the two models
(with and without a cosmological constant) 
start to deviate significantly from one another.
However, it is worth noting that
the faster evolution in the low density models than
in the high density models is somewhat coincidental (due to $\sigma_8-\Omega_0$
relation required to fit local rich cluster abundance, equation 36 of C98)
and entirely due to the normalization on $\sigma_8$,
not due to $\Omega_0$, as will be shown below.

\begin{figure*}
\centering
\begin{picture}(400,250)
\psfig{figure=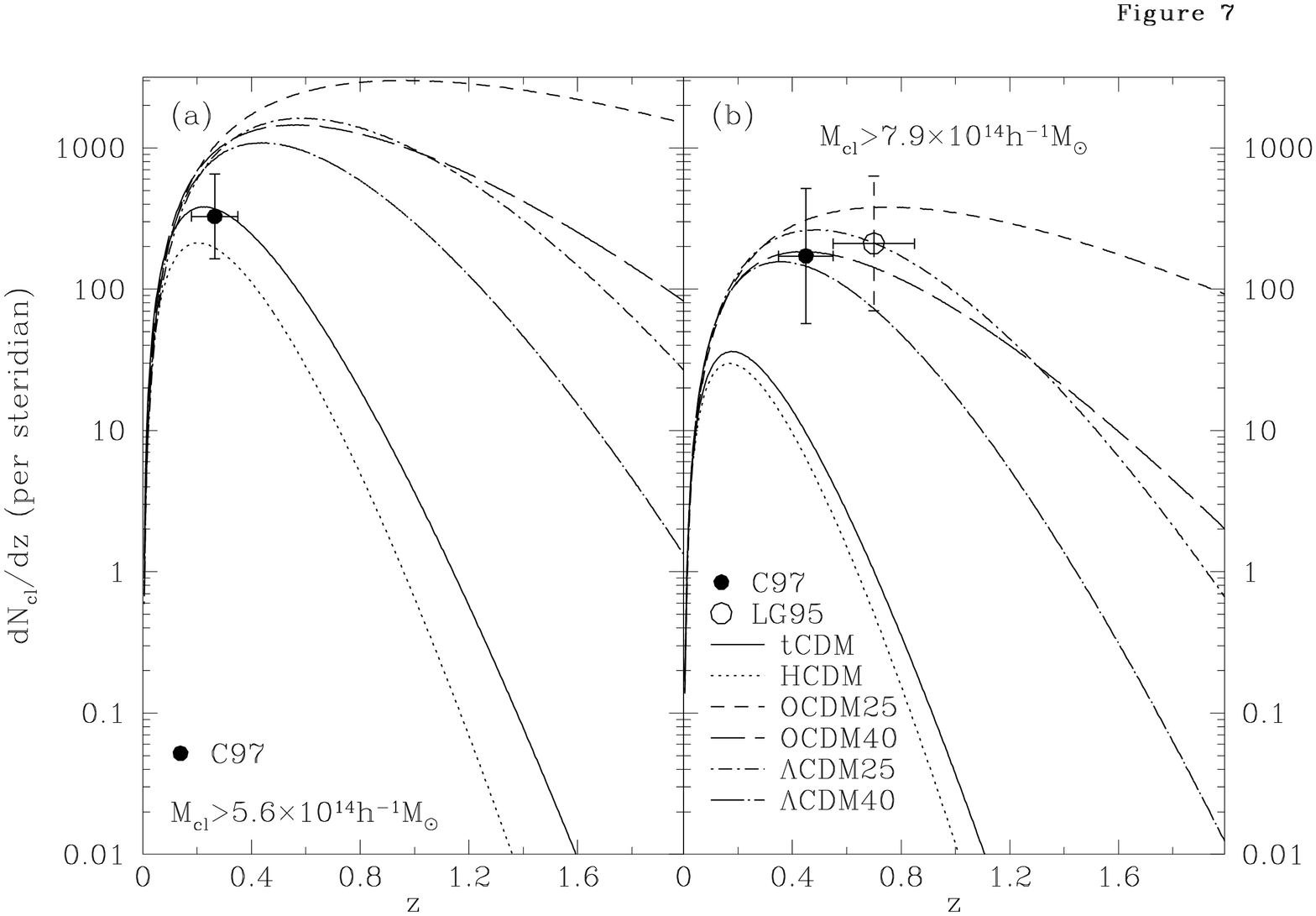,height=10.0cm,width=15.0cm,angle=270.0}
\end{picture}
\caption{
Panel (a) shows the number of
clusters with mass $\ge 5.6\times 10^{14}h^{-1}\msun$
within a fixed {\it proper} radius of $1.5h^{-1}$Mpc, and
panel (b) shows the evolution of the number of
clusters with mass $\ge 7.9\times 10^{14}h^{-1}\msun$
within a fixed {\it proper} radius of $1.5h^{-1}$Mpc.
The quantity in the vertical axis
is the number of clusters per unit redshift per steradian.
Shown as symbols are the observations with $2\sigma$ errorbars
(Carlberg \etal 1997, C97; Luppino \& Gioia 1995, LG95).
}
\end{figure*}

Presenting the above results shown 
in Figures (5,6) in a slightly different form,
panel (a) of Figure 7 shows the evolution of the number of 
clusters with mass of $\ge 5.6\times 10^{14}h^{-1}\msun$
within a fixed {\it proper} radius of $1.5h^{-1}$Mpc,
and panel (b) of 
clusters with mass of $\ge 7.9\times 10^{14}h^{-1}\msun$.
The quantity in the ordinate
is the number of clusters per unit redshift per unit steradian.
Since the 
number of clusters per unit redshift {\it per unit sold angle} is plotted,
one has to 
take account the different geometries in different model universes
(the comoving volume element per steradian per unit redshift
about $z$ is $D_A(z)^2 dr_{cm}(z)/dz$, where
$D_A$ is the comoving angular diameter distance to redshift $z$
and $r_{cm}(z)$ is the comoving distance to redshift $z$;
see equations 6, 8 below for definitions of $D_A$ and
$r_{cm}$).
It is seen 
that the less massive, low redshift clusters
shown in panel (a) has much less
discriminating power than the more massive, higher
redshift clusters shown in panel (b).
While the observed clusters at $z=0.18-0.35$ (panel a) are not able to 
differentiate between any models at a significant confidence level,
the observed clusters at $z=0.35-0.55$ and $z=0.55-0.85$ (panel b)
offer a clean way to separate the $\Omega_0=1$ models 
from the rest.
It seems quite secure to conclude that
both $\Omega_0=1$ tCDM and HCDM models are ruled out at a high
confidence level ($>3\sigma$),
whereas the current observations are not able to differentiate
between the four low density models.
This conclusion is in broad agreement with the conclusions
drawn by Henry \& Arnaud (1991), Bahcall, Fan \& Cen (1997)
and Carlberg \etal (1997).
Reichart \etal (1998) claim that the X-ray cluster 
luminosity function from Einstein Medium Sensitivity Survey
are consistent with $\Omega_0=1$.
It is unclear at this time what causes the difference.
However, Reichart \etal (1998) 
 claim that their conclusion
 is independent of $\sigma_8$ (i.e., the amplitude
of the power spectrum), which is in direct conflict with 
the demonstration that $\sigma_8$ is the primary factor
that determines the redshift evolution of massive clusters,
as shown below.

\begin{deluxetable}{cccccccccc} %{l,r}
\tablewidth{0pt}
\tablenum{2}
\tablecolumns{4}
\tablecaption{Number of High Redshift X-ray Clusters\tablenotemark{a}~~in Six Models} %\label{tab1}}
\tablehead{
\colhead{Model} &
\colhead{\vbox{\hbox{(5.0,7.0)keV}\hbox{\ \ $z>0.5$}}} &
\colhead{\vbox{\hbox{(5.0,7.0)keV}\hbox{\ \ $z>1.0$}}} &
\colhead{\vbox{\hbox{(5.0,7.0)keV}\hbox{\ \ $z>2.0$}}}}

\startdata
tCDM & $(71,2.2)$ & $(1.3,0.01)$ & $(0.0,0.0)$ \nl 
HCDM & $(24,1.3)$ & $(0.19,0.0028)$ & $(0.0,0.0)$ \nl 
OCDM25 & $(14902,1374)$ & $(10591,798)$ & $(3232,139)$ \nl 
OCDM40 & $(3323,273)$ & $(1314,70)$ & $(75,1.4)$ \nl 
$\Lambda$CDM25 & $(3217,369)$ & $(1005,72)$ & $(16,0.30)$ \nl 
$\Lambda$CDM40 & $(1270,112)$ & $(203,9.1)$ & $(0.56,0.0040)$ \nl 
\enddata
\tablenotetext{a}{for a sky coverage of $\pi$ steradian}
\end{deluxetable}

To quantify the power of high redshift X-ray 
clusters for differentiating between models,
we tabulate in Table 2 the expected
number of clusters in each of the six models 
at redshifts greater than 0.5, 1.0 and 2.0, respectively,
for clusters with temperatures greater than 0.5keV and 7.0keV, 
respectively.
The numbers are obtained assuming that 
a survey covering  $\pi$ steradian (i.e., one quarter of the sky)
is available.
Next generation X-ray telescopes should be able to 
detect $7keV$ clusters at $z=1$ without much difficulty,
if clusters at high redshift are {\it not significantly}
less luminous than their local counterparts at
a fixed temperature (e.g., Henry 1997).
It is clear from Table 2
that $\Omega_0=1$ models
are easily differentiated from other models.
The subtlety lies in low density models and some degeneracy
exists between open models and flat models with cosmological
constants.
For example, OCDM40 ($\Omega_0=4.0$, $\Lambda_0=0.0$)
and $\Lambda$CDM25 ($\Omega_0=0.25$, $\Lambda_0=0.75$)
give comparable number of bright X-ray clusters at $z<1.0$;
but the difference is gradually larger at higher redshift
(a factor about four at $z=2$).
Aside from this degeneracy, the database from a survey
of the indicated sky coverage should be
able to determine $\Omega_0$ with an uncertainty
of $\Delta\Omega_0=0.02$ ($1\sigma$),
if $\Omega_0+\Lambda_0=1$
[obtained by assuming  statistical (Poisson) errors
for the clusters with $kT=7$keV at $z>1$
of the last two models in Table 2, which exhibit
the smallest difference and thus are most demanding].
If the universe is open, one will be able to 
determine $\Omega_0$ with an accuracy of 
$\Delta\Omega_0=0.008$ (by interpolating between OCDM25 and OCDM40).
These accuracies are comparable to those
which will be determined from the next generation
microwave background fluctuation experiments 
(e.g., Zaldarriaga, Spergel, \& Seljak 1997),
thus provide a very powerful test and also
a very important consistency cross-check.

\begin{figure*}
\centering
\begin{picture}(400,300)
\psfig{figure=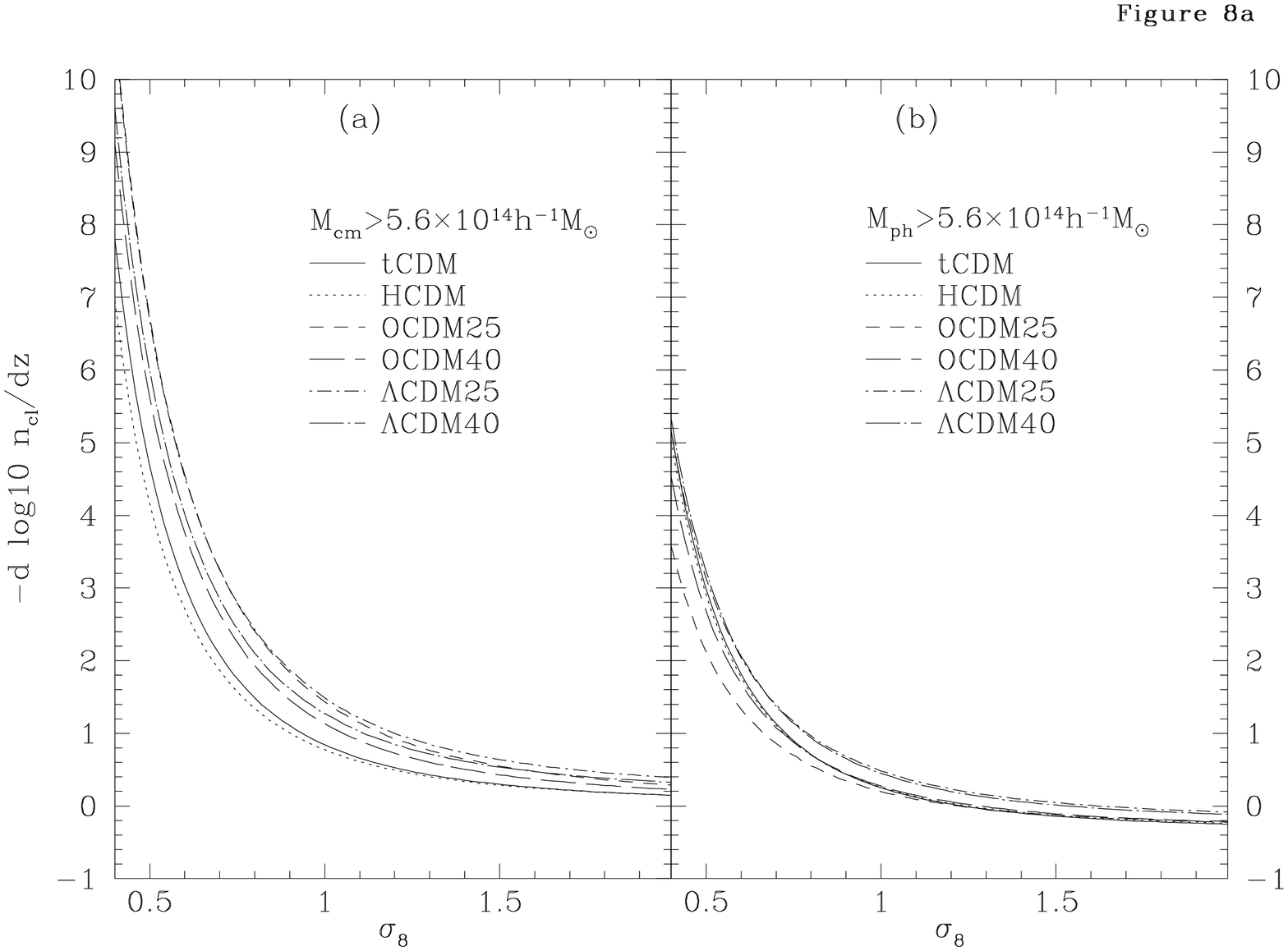,height=10.0cm,width=15.0cm,angle=270.0}
\end{picture}
\centerline{(8a)}\vspace{0.1in}
\end{figure*}

\begin{figure*}
\centering
\begin{picture}(400,300)
\psfig{figure=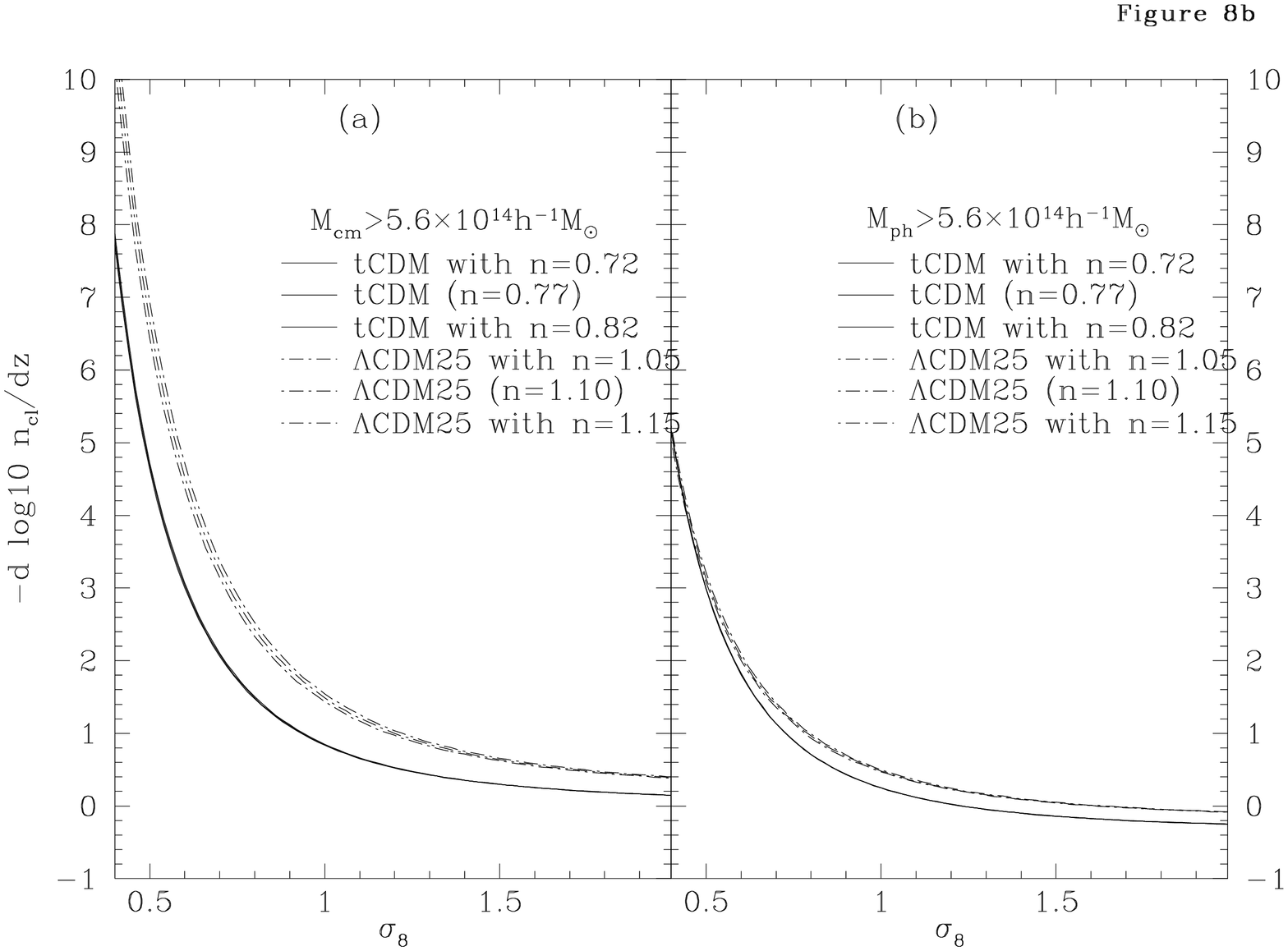,height=10.0cm,width=15.0cm,angle=270.0}
\end{picture}
\centerline{(8b)}\vspace{0.1in}
\caption{
(8a) shows the evolution rate of cluster
number density 
as a function of the amplitude
of the power spectrum ($\sigma_8$)
for the six models (Table 1).
Panels (a,b) of Figure (8a)
show the evolution rate of number density for clusters
whose mass (=$5.6\times 10^{14}h^{-1}\msun$)
is defined within a comoving Abell radius and within
a physical Abell radius, respectively.
(8b) shows
the dependence of the evolution rate on power index $n$
by varying $n$ for tCDM and $\Lambda$CDM25.
}
\end{figure*}

So far we have shown that there are dramatic differences
in the number density of massive clusters at high redshift
between the considered models,
when all are normalized to give the same cluster density at $z=0$.
In other words, the rate of evolution for 
cluster abundance is strongly model-dependent.
It is useful to understand what cause such dramatic differences
in the evolution rates between the different models.
For this let us first define an evolution rate of cluster
(with mass $>M$)
number density as a function of $\sigma_8$ as
\begin{equation}
{\cal E}(\sigma_8, >M) \equiv -\frac{d\log_{10}n(\sigma_8, >M)}{dz}.
\end{equation}
\noindent 
This is the change of cluster abundance in logarithmic scale per unit redshift.
This definition is similar to 
that defined in Fan, Bahcall, \& Cen (1998; equation 8)
but with the difference that the current ${\cal E}$ is a 
function of $\sigma_8$ and $M$, not a function of $z$ and $M$.
In this sense the current definition 
of rate of evolution is local in time;   
redshift $z$ in equation 2 serves as a ruler or a unit.
Figure 8a shows the evolution rate of the
number density of clusters
with mass $>5.6\times 10^{14}h^{-1}\msun$,
defined within a comoving Abell radius (panel a) and within
a physical Abell radius (panel b), respectively.
The first thing to notice is that, not surprisingly,
the evolution rate is lower for clusters defined within a
physical Abell radius than within a comoving Abell radius.
Secondly, 
{\it the differences between models at a fixed $\sigma_8$ 
are smaller than the differences of any single model at different
$\sigma_8$'s.}
For example, for OCDM25 
${\cal E}$ changes from $1.4$ to $6.8$
for $\sigma_8$ from $1.0$ to $0.5$,
while the differences in ${\cal E}$ between tCDM and OCDM25 
is $0.6$ at $\sigma_8=1.0$ and $2.0$ at $\sigma_8=0.5$.
However, it is interesting to note that,
{\it at a fixed $\sigma_8$},
the evolution rates in the low $\Omega_0$ models 
are {\it higher, not lower}, 
than the rates in the $\Omega_0=1$ models
for clusters defined within the comoving Abell radius (panels a).
The situation is altered for clusters defined within
the physical Abell radius: 
{\it at a fixed $\sigma_8$},
the evolution rate 
is {\it lowest} in OCDM25.
In any case these differences are secondary compared to the
difference of $\cal E$ between different $\sigma_8$ within a same model.
Next, we select two models (tCDM and $\Lambda$CDM25)
and let the power index $n$ vary within the allowable ranges
to test the dependence of ${\cal E}$ on $n$.
The results are shown in Figure 8b, from
which we can easily draw the conclusion
that the dependence of the evolution rate on $n$ is very weak.

We conclude that 
the dramatic differences
in the evolution rates between the different models
are {\it almost entirely caused by the differences in $\sigma_8$}
between the models, which results from the 
requirement that each model has to fit
the cluster abundance at $z=0$.
{\it The apparent dependence of the evolution rates on $\Omega_0$/$\Lambda_0$
is dictated by the strong dependence of cluster abundance
on $\sigma_8$ through the $\sigma_8-\Omega_0$ relation}
(equation 36 of C98).
This conclusion is in good agreement with 
that of Fan \etal (1998), who use the Press-Schechter method.

\subsection{Gravitational Lensing by Clusters of Galaxies}

Clusters of galaxies are so massive
that they ubiquitously cause noticeable distortions
of images of background 
sources even in their outskirts ($\sim 1~h^{-1}$Mpc)
(\cite{tvw90}; \cite{m91}; \cite{ks93}).
The central regions of rich clusters are often capable of 
producing more spectacular events such as 
large splitting, multiple images (of background quasars)
or giant arcs (of background galaxies) 
(e.g., Le Fevre \etal 1994;
Edge \etal 1994; Smail \etal 1996).
These extreme gravitational lensing events are rare
because the cross section
for such events is small; usually only the central region
of size of order several kiloparsecs in a very rich
cluster is capable of
producing giant arcs or multiple splittings of size of  
arcseconds and larger,
compared to the usual notion of the size of a cluster ($\sim 1$Mpc)
or even the size of the X-ray core ($\sim 100$kpc) of a cluster.
Consequently, the frequency of such events
is extremely sensitive to abundance of a very small
population of very rich clusters at intermediate redshifts
($\sim 0.3-2.0$), where lensing is optimal.

Narayan \& White (1998) compare strong lensing events in various
models (pre-COBE) with observations combining Press-Schechter 
method with the singular isothermal sphere assumption.
Cen \etal (1994) and Wambsganss \etal (1995) use detailed
N-body simulations to compute the lensing statistics,
focusing on intermediate splittings (a few arcseconds).
Here we compute the frequency of very large splitting 
or giant arc events ($\sim arcminute$).
We approximate each cluster as a singular isothermal sphere (SIS;
Turner, Ostriker \& Gott 1984, TOG henceforth)
with the one-dimensional
velocity dispersion determined by the mass within
the comoving Abell radius:
\begin{equation}
\sigma_{||}^2={G M_A (1+z_L)\over 2 r_A},
\end{equation}
\noindent 
where $G$ is the gravitational constant,
and $z_L$ is the redshift of the (lens) cluster in question.
For a SIS the bending angle, $\theta_0$, 
conveniently independent of impact parameter,
is (TOG)
\begin{equation}
\theta_0= {4\pi}({\sigma_{||}\over c})^2,
\end{equation}
\noindent 
where $c$ is the speed of light. 
The size of the arc due to such a lens
will be approximated as the diameter of 
the Einstein ring as:
\begin{equation}
\beta= {2\theta_0 D_{LS}\over D_S},
\end{equation}
\noindent 
where $D_{LS}$ and $D_S$ are the (comoving) angular  diameter distance
between the lens (cluster) and the source, and
between observer and the source, respectively.
The (comoving) angular diameter distances are
dependent upon the geometry of a world model 
[$D_{LS}=D(\chi_S-\chi_L)$ and $D_S(\chi_S)$] as:
\begin{equation}
{D(\chi)=\cases{&\hskip -0.5cm$a_0\chi\hskip 1.44cm\hbox{for flat universe}\quad (\Omega_0+\Lambda_0=1)$ \cr 
&\hskip -0.5cm$a_0\sin (\chi)\hskip 0.62cm\hbox{for close universe}\quad (\Omega_0+\Lambda_0>1)$\cr
&\hskip -0.5cm$a_0\sinh (\chi)\hskip 0.4cm\hbox{for open universe}\quad (\Omega_0 + \Lambda_0<1)$\cr}}
\end{equation}
\noindent 
Here, $a_0$ is a dimensional scale factor defined below;
$\chi$ is the dimensionless comoving distance to the object
in question at $z$, which can be expressed for an 
arbitrary universe 
(Refsdal, Stabell \& de Lange 1967; Gott, Park \& Lee 1989) as 
\begin{equation}
\chi(z) = \sqrt{|3\sigma_0-q_0-1|}\int_0^z Z(u)du,
\end{equation}
\noindent 
where $Z(u) = [1+2(q_0+1)u +(q_0+1+3\sigma_0)u^2+2\sigma_0u^3]^{-1/2}$.
Here $\sigma_0=\Omega_0/2$ and 
$q_0$ is the deceleration parameter, 
which is $q_0=\Omega_0/2$ for $\Lambda_0=0$ and
$q_0=3\Omega_0/2 - 1$ for $\Omega_0+\Lambda_0=1$.
The dimensional scale factor $a_0$ is 
$a_0 = {c\over H_0 \sqrt{|3\sigma_0-q_0-1|}}$,
where $H_0$ is the Hubble constant.
The dimensional comoving distance of an object at $z$
is 
\begin{equation}
r_{cm}(z) = a_0 \chi (z).
\end{equation}
\noindent 
Combining equations 3,4,5, we obtain an expression for 
the required cluster mass $M_A$ in order to produce 
an arc of size $\beta$ as
\begin{equation}
M_A(\beta)= {\beta c^2 r_A \over 4\pi G(1+z_L)}{D_{S}\over D_{LS}}.
\end{equation}
\noindent 
We can then compute the optical depth per unit redshift for 
forming arcs of size $>\beta$, as a function of redshift:
\begin{equation}
{d\tau\over dz} = \pi \beta^2{dN_{cl}\over dz},
\end{equation}
\noindent 
where ${dN_{cl}\over dz}$ is the number of clusters with
mass $>M_A(\beta)$ (equation 9)
per unit steradian per unit redshift,
computed using GPM (C98).

\begin{figure*}
\centering
\begin{picture}(400,300)
\psfig{figure=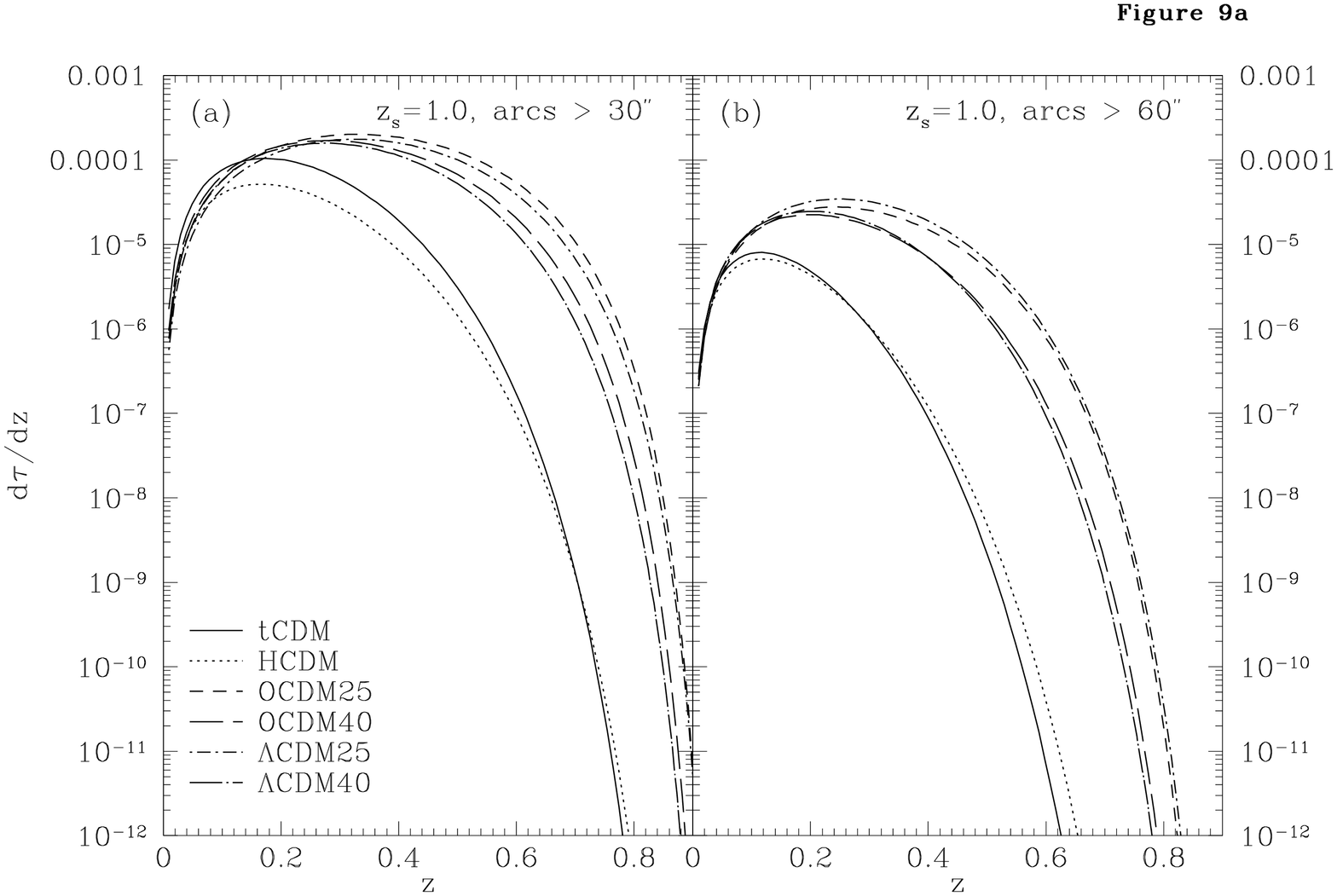,height=10.0cm,width=15.0cm,angle=270.0}
\end{picture}
\centerline{(9a)}\vspace{0.1in}
\end{figure*}

\begin{figure*}
\centering
\begin{picture}(400,300)
\psfig{figure=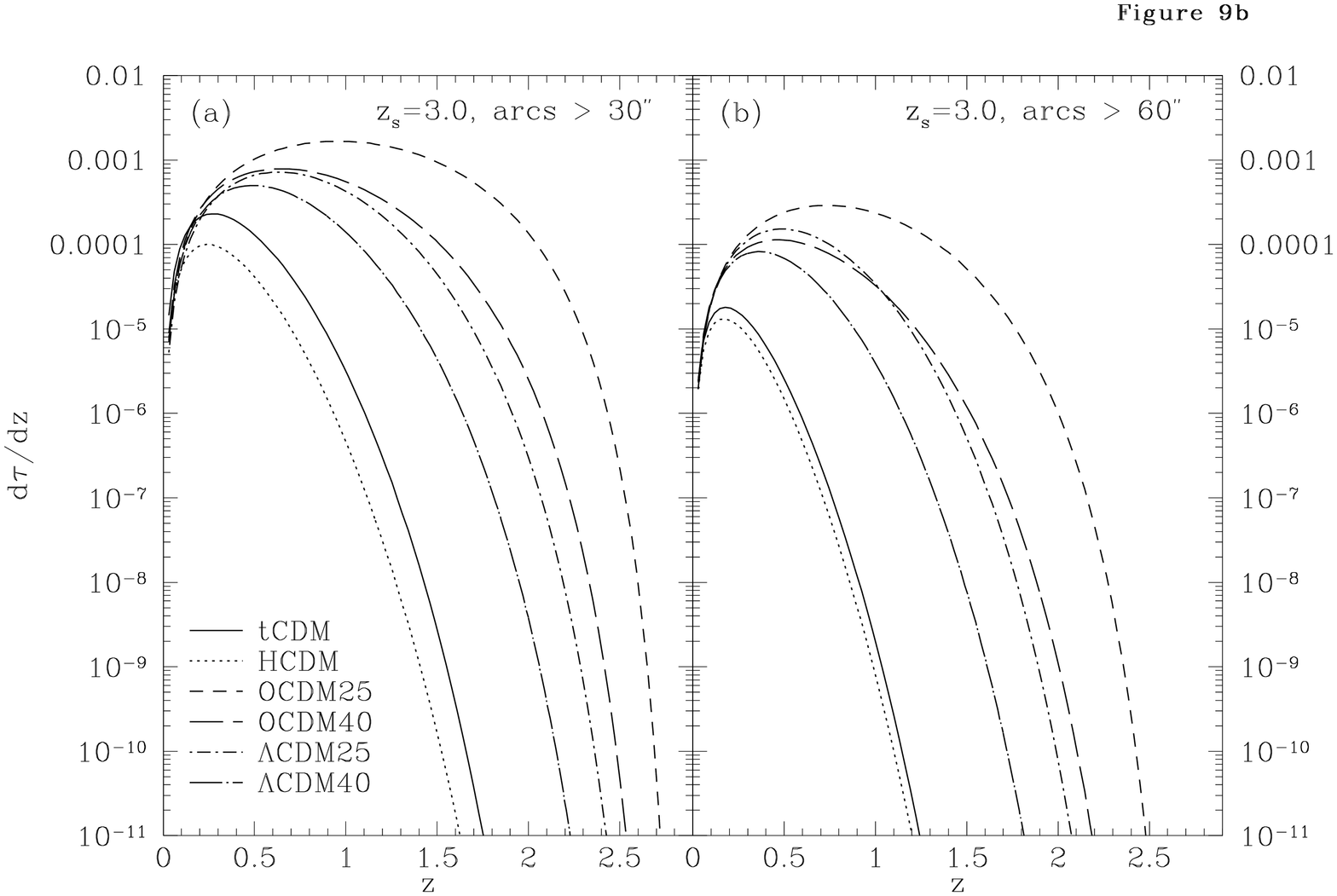,height=10.0cm,width=15.0cm,angle=270.0}
\end{picture}
\centerline{(9b)}\vspace{0.1in}
\caption{
The differential
gravitational lensing
probability (i.e., differential optical depth)
per unit redshift as a function of redshift
for arcs or splitting angles of multiple image events
with sizes greater than
$30^"$ and $60^"$ for sources at $z=1$ (9a) and $z=3$ (9b), respectively.
}
\end{figure*}

Figures (9a,b) show the (differential) gravitational lensing
probability (i.e, differential optical depth)
per unit redshift as a function of redshift
for arcs or splittings greater than
$30^"$ and $60^"$ for sources at $z_s=1$ and $z_s=3$, respectively.
In accord with the evolution of cluster abundance
as shown in Figures (5,6,7),
the $\Omega_0=1$ models (tCDM, HCDM)
have the lowest optical depths due to the rapid
decrease in the number of rich cluster towards high redshift.
The lowest $\Omega_0$ models (OCDM25, $\Lambda$CDM25) have the 
highest optical depths.
However, the difference between OCDM25 (open $\Omega_0=0.25$)
and $\Lambda$CDM25 (flat $\Omega_0=0.25$) 
(similarly between OCDM40 and $\Lambda$CDM40)
is more subtle:
for sources at $z_s=1$ $\Lambda$CDM25 has a somewhat higher
optical depth than OCDM25 because the 
prolonged path length due to the cosmological constant
in $\Lambda$CDM25 plays a significant role while the evolution
rate of cluster abundance in the two models are comparable out to 
redshift $\sim 0.5$;
but for sources at $z_s=3$ OCDM25 has a significantly higher
optical depth than $\Lambda$CDM25 because the much lower rate of 
cluster abundance evolution in OCDM25 dominates over
other factors.

The assumption of SIS is admittedly approximate,
but used because of its analytical simplicity 
and because of its astrophysical relevance 
as used in many statistical studies 
(e.g., Tyson 1983; TOG;
Narayan \& White 1988;
Fukugita \& Turner 1991;
Mao 1991).
A perhaps more realistic profile is 
NFW profile (Navarro \etal 1996), which goes as $r^{-1}$ in the center
and as $r^{-3}$ in the outskirt.
Which profile, NFW or SIS, is more 
efficient for lensing depends on the core radius of clusters. 
If the core radius is sufficiently small, the NFW profile could be
more efficient than a SIS profile, for a given cluster mass
within a larger, given radius (Maoz \etal 1997).
On the other hand, if the SIS assumption overestimates
the strong lensing probability,
this would diminish the already 
very small lensing optical depth
in the high $\Omega_0$ models further (the reductions in model
with higher optical depths are smaller),
widening the difference between models.
Given the simplicity of our treatment,
we choose to focus on the 
gross frequency of large gravitational arcs 
or large splitting multiple images,
which are relatively robust,
{\it not on the details such as 
length-to-width ratio of arcs}.
The latter may sensitively depend on the small scale
structures in the clusters such as core radius, inner density 
slope and substructure
(see Bartelmann \etal 1997 for an excellent study on arcs 
of smaller sizes).
The lack of or inadequate treatment of galaxy formation
and inaccuracy in the small scale density distributions
(even with very large, pure N-body calculations currently available)
do not yet permit a sufficiently accurate characterization of 
lens potentials on small scales.

Clearly, the gravitational
lensing optical depths of large arcs (or large splitting images)
for different models 
deviate rapidly from one another towards high redshift.
We argue that the errors introduced by our
approximate method are likely to be substantially smaller than
the difference between the models at moderate-to-high redshift.
Detection of even a small sample (or non-detection) of such large lenses 
at redshift $>1.0$ 
and/or a large sample of high redshift sources with such 
large arcs will provide a useful constraint on the models.
Systematic surveys for large-image separation lensed quasars 
(Maoz \etal 1997) or arcs
are beginning to provide useful statistical samples,
and larger data sets such as from SDSS will be extremely
powerful in constraining cosmological models.
With the data set of $10^6$ quasars 
from SDSS
(assuming to be 
at $z_s=3$ for this simple illustration),
we expect to see (inferred from panel a of Figure 9b)
($1900\pm 44$, $710\pm 27$, $570\pm 24$, $330\pm 18$, $100\pm 10$, $39\pm 6$)
multiple-image quasars  with image separation $\ge 30^"$
in (OCDM25, OCDM40, $\Lambda$CDM25, $\Lambda$CDM40, tCDM, HCDM),
respectively (where the uncertainties are $1 \sigma$ Poissonian).
This would allow an unambiguous test of these six models.
More likely, it would put a constraint on 
$\Omega_0$ with $\Delta\Omega_0\sim 0.01$ if the
universe is open or 
$\Delta\Omega_0\sim 0.03$ if the
universe is flat.
A more stringent test can be obtained
if one also has the redshift distribution of the lenses.
One expect to see 
(200.0, 4.0, 0.5, 0.005, 0.0, 0.0) lenses
at $z\ge 2$ in
(OCDM25, OCDM40, $\Lambda$CDM25, $\Lambda$CDM40, tCDM, HCDM),
respectively, that produce arcs or splittings of size $\ge 30^"$.
We see that the differences between models in this measure
are even more dramatic.

\subsection{$\lya$ Forest}
 
The abundant low column density $\lya$ forest lines in quasar
absorption lines provide a precise and unbiased
tracing of the distributions of neutral hydrogen,
in a variety of environments,
up to the highest redshift where quasars are observed to exist
(e.g., Rauch \etal 1992; Schneider \etal 1993; Hu \etal 1995;
Tytler \etal 1995; 
Lanzetta \etal 1995a; Bahcall \etal 1996).
Until very recently, models of quasar absorption systems 
were {\it post hoc}.
Successfully putting the formation and evolution of $\lya$
clouds into a cosmological context, computed {\it ab initio}
from standard cosmological initial conditions
(Cen \etal 1994; Zhang \etal 1995;
	Hernquist \etal 1996;
Miralda-Escud\'e \etal 1996, MCOR hereafter; Bond \& Wadsley 1997),
has set forth a brand new way to study these objects.
The surprisingly good
agreement between simulations and observations
in terms of matching the distributions of line width parameter $b$
and column density, among other quantities,
clearly demonstrates that the basic cosmological 
origin for $\lya$ clouds is no longer questionable.

\begin{figure*}
\centering
\begin{picture}(400,260)
\psfig{figure=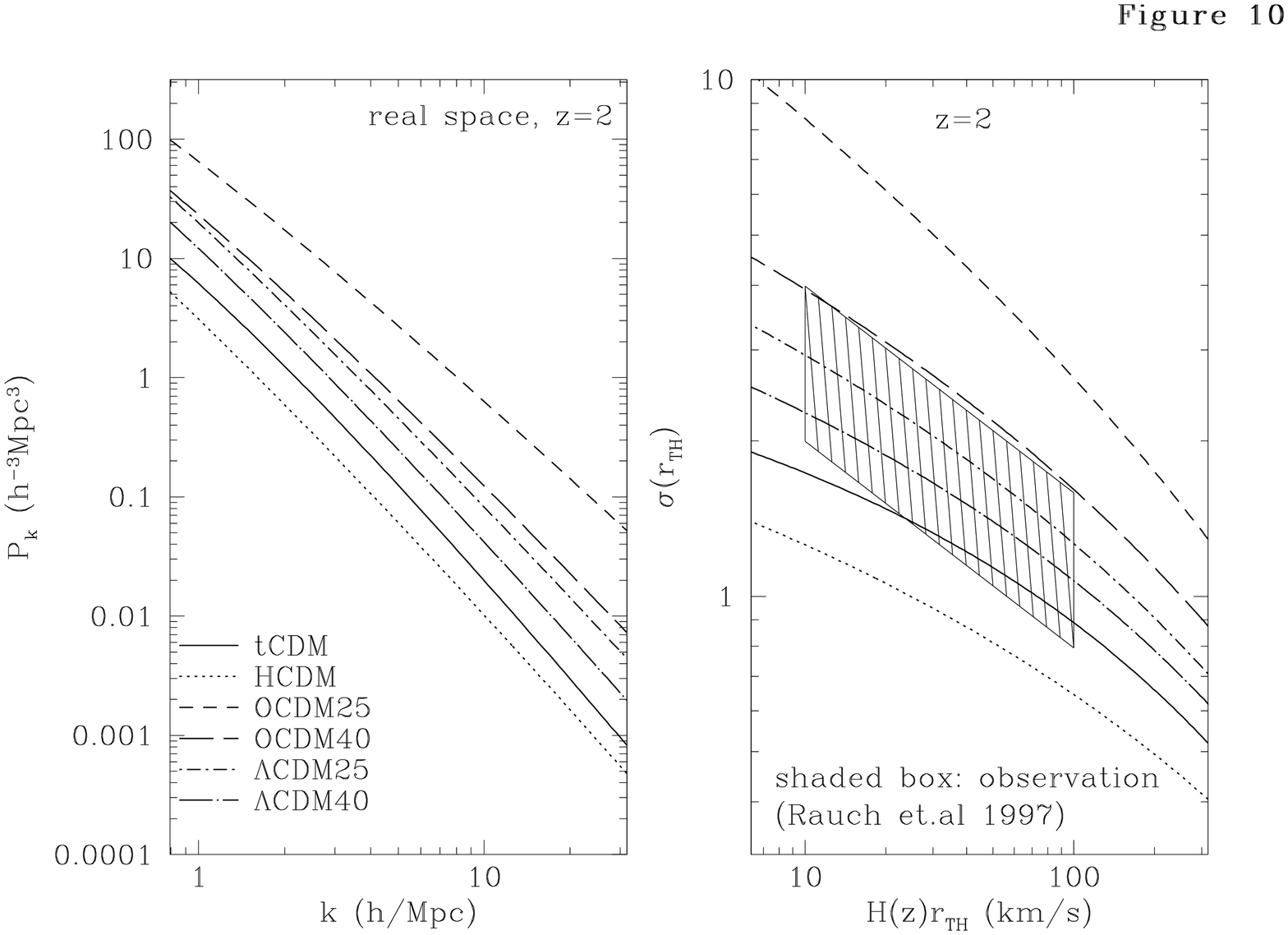,height=10.0cm,width=15.0cm,angle=270.0}
\end{picture}
\caption{
The power spectra (panel a)
and the rms density fluctuations (panel b), both in real space,
at the relevant scales ($k\sim 1.0-30.0h/$Mpc and $H(z) r_{TH}\sim 10.0-300.0$km/s) at $z=2$, computed using linear theory.
Also shown as a cross-shaded region
is the observational constraint adapted from Rauch \etal (1997).
In panel (b) the x-axis is in km/s, obtained by multiplying
the physical scale ($r_{TH}$) by the Hubble constant at $z=2$, $H(z)$.
}
\end{figure*}

It is to distinguish between competing cosmological models
that interests us most here.
We attempt to check illustratively
if $\lya$ forest may be used to test cosmological models.
We will only examine models at $z=2$, where good observational
data are available.
Figure 10 shows the power spectra (panel a)
and the rms density fluctuations (panel b), both in real space
and computed using linear theory,
at the relevant scales ($k\sim 1.0-30.0h/$Mpc and 
$H(z) r_{TH} \sim 10.0-300.0$km/s) at $z=2$. 
In order for the models 
to be shown and compared on a common, relevant physical scale,
the Hubble velocity instead of proper (or comoving)
physical distance is shown in the x-axis in panel (b).
Also shown as a cross-shaded region
is the observational constraint adapted from Rauch \etal (1997).
The adopted observational constraint is based on
the range of the fluctuation amplitudes on the relevant scales
spanned by the two models (SCDM and $\Lambda$CDM; 
Figure 8 of Rauch \etal 1997)
examined by Rauch \etal (1997), which fit the observations well
and provide a useful indication in the needed power on the relevant scales.
The plotted vertical range of the shaded box is three times
the range between SCDM and $\Lambda$CDM in Rauch \etal (1997),
centered on the mean amplitude (at each x-axis value)
averaged over the SCDM and $\Lambda$CDM models.

The disparity of powers between the models on these scales relevant to
the formation of $\lya$ clouds is rather striking. 
The very slow growth of perturbations at low density open models
is clearly manifested:
OCDM25 (open $\Omega_0=0.25$)
has distinctly more power on all the shown scales than
other models, 
while HCDM 
has power at $k=30h/$Mpc lower than that of OCDM25 by almost two
orders of magnitude.
These differences 
should result in distinctly
different populations of $\lya$ forest lines and flux distributions.
We expect that the overall flux distributions 
(Rauch \etal 1997) of the models will be different,
even with the liberty to adjust the ionization rate to fit
the observed mean flux decrement
(Zuo \& Lu 1993; \cite{prs93}).
In particular, the low flux end 
of the 1-point flux distribution will be
dramatically different between the models (Cen 1997a).
Since the amount of observational data available
(and accumulating) on $\lya$ forest lines are very large,
measures based on flux distributions (MCOR; Cen 1997a;
Miralda-Escud\'e \etal 1997)
may enable tests of models with high statistical significance. 
Preliminary comparisons between models and observations
using the one-point flux distribution probability function
(Rauch \etal 1997)
seem to most favor the two $\Lambda$ based models 
($\Lambda$CDM25 and $\Lambda$CDM40),
but disfavor HCDM and OCDM25 in the opposite sense that
HCDM has too little power and OCDM25 has too much power
on the relevant scales.
tCDM and OCDM40 appear to be marginally consistent.

Definitive theoretical predictions of all the models
and comparisons between models and observations
await detailed modeling using 
cosmological hydrodynamic simulations
suitable to study the formation of $\lya$ clouds (Cen \etal 1994;
Zhang \etal 1995).

\subsection{Damped $\lya$ Systems}

The high end of the $\lya$ clouds 
is the damped $\lya$ systems,
which are thought to be progenitors of large disk galaxies (Wolfe 1995).
Damped $\lya$ systems are much 
rarer than the common $\lya$ forest lines
by approximately two orders of magnitude,
analogous to relative abundances between galaxies and clusters of galaxies
at present time.
In this sense damped $\lya$ systems 
probe the tail of the density distribution
on $\sim 1h^{-1}$Mpc scale
at high redshift,
whereas present clusters of galaxies 
probe the tail of the density distribution
on $\sim 10h^{-1}$Mpc scale.

\begin{figure*}
\centering
\begin{picture}(400,270)
\psfig{figure=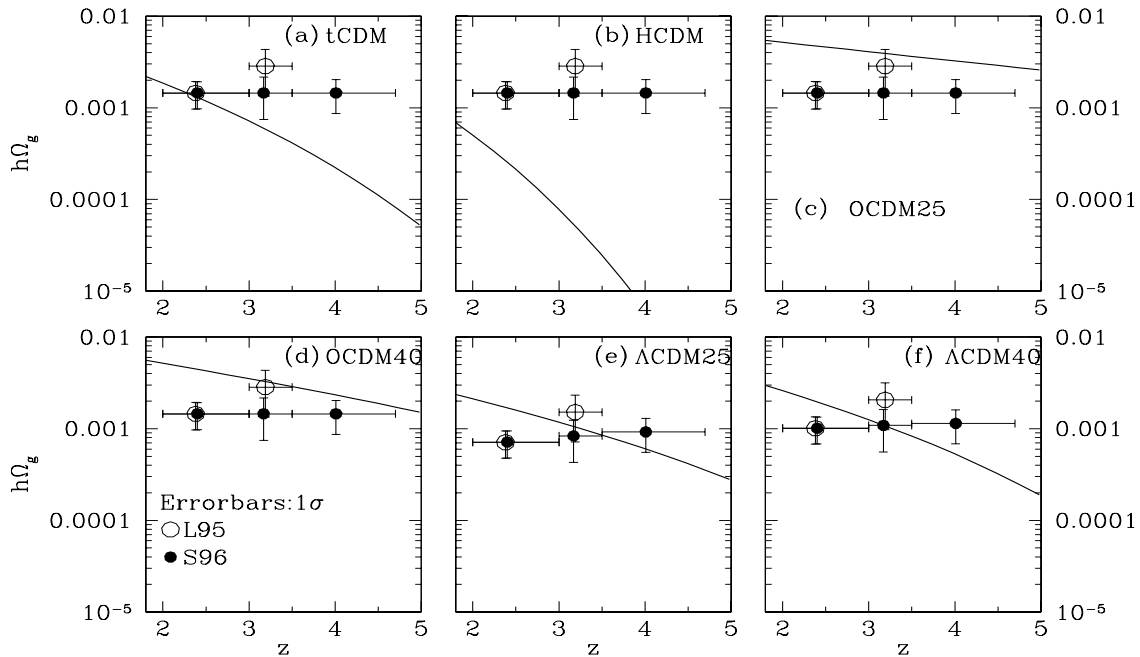,height=10.0cm,width=15.0cm,angle=270.0}
\end{picture}
\caption{
The gas density of collapsed baryons in systems
with rotation speed of $>180\kms$,
as a function of redshift for six models.
The observations are from Lanzetta \etal (1995)
and Storrie-Lombardi \etal 91996) with $1\sigma$ errorbars.
}
\end{figure*}

Prochaska \& Wolfe (1997) find that 
the only model among tested 
that matches the observed kinematics
of damped $\lya$ systems is
a thick fast rotating disk of
rotation speed of $>180\kms$ (99\% confidence level)
with 
the most likely rotation velocity being $225$km/s.
This thus seems to provide an additional
piece of evidence that 
damped $\lya$ systems may be the progenitors of present 
large disk galaxies.
We adopt the observed lower bound on the rotation
speed of $180\kms$ to compare with models.
Figure 11 shows the density of collapsed baryons in systems
with rotation velocity 
(equation 18 of C98)
greater than $180\kms$ as a function
of redshift,
computed using GPM (C98).
The observations are from Lanzetta \etal (1995b, L95 henceforth and
in Figure 11)
and Storrie-Lombardi \etal 91996, S96 henceforth) with $1\sigma$ errorbars.
The errorbars are those adopted in S96 for both sets of observations.
A small complication arises from the fact
that the observations are tabulated only for 
for models with $q_0=0$ and $q_0=0.5$
[Table 1 of S96].
We compute the observed values for other models
by linear interpolating $h \Omega_g$ between the two tabulated models
using weighting of the inverse of the 
Hubble constant at each redshift bin.
This is justified since the absorption
distance interval is inversely proportional
to the Hubble constant at the redshift in question:
$\Delta X=(1+z)^2 [H_0/H(z)]$ (Bahcall \& Peebles 1969),
where $H(z) = H_0\sqrt{\Omega_0(1+z)^3+(1-\Omega_0-\Lambda_0)(1+z)^2 + \Lambda_0}$ (Peebles 1993).
It is, however, still approximate due to 
non-uniform distributions
of absorption redshift paths within each bin
and finite sizes of the redshift bins.

Without running detailed hydrodynamic simulations 
we wish only to {\it conservatively}
assume that all the baryons (which is proportional
to the mass with a fixed ratio equal to the global ratio)
that collapse into these systems are in neutral form.
This is conservative in the sense that it is an upper
bound on $\Omega_g$, since some amount of hydrogen may be
in other forms such as ionized or molecular form.
Bearing in mind that the computed quantity
is an upper bound,
we see that HCDM is inconsistent
with observations at a high confidence level, 
in agreement with the conclusion drawn by Ma \etal (1997) for a
comparable model,
and consistent with previous analyses
(Subramanian \& Padmanabhan 1994;
Kauffmann \& Charlot 1994;
Mo \& Miralda-Escud\'e 1994;
Klypin \etal 1995),
which use an HCDM model with a somewhat larger fraction 
of mass being in hot neutrinos ($\Omega_h=0.3$).
Note that the newer observation with significantly absorption 
path length in S96 than in L95
has revised and significantly lowered the point at $z=3.19$.
But the basic conclusions previously reached by other authors 
with regard to HCDM model remain valid.
All the other models are consistent with observations with
perhaps the exception of OCDM25 (open $\Omega_0=0.25$),
which appears to have overproduced the neutral hydrogen gas
in these systems.
But we recall that photoionization and
other heating sources could significantly reduce
the neutral hydrogen fraction in these systems.
Therefore, OCDM25 cannot be ruled out yet without
more detailed calculations.

\subsection{High Redshift Galaxies}

Recently, observations of high redshift galaxies have
made major break-throughs by using an innovative technique
to systematically and efficiently
search for them (Pettini \etal 1997; Steidel \etal 1998, S98 henceforth).
Interesting discoveries 
have already been made with a relatively small sample
of such galaxies ($\sim 200$).
S98 find a large concentration of galaxies
in redshift space near redshift three in a field of size
about $9\times 15~$arcmin$^2$.
The concentration is so large for $z\sim 3$
that S98 are able to 
draw some interesting implications for cosmological models and 
galaxy formation.
We will use a somewhat different method to 
conservatively assess the implications of such high
redshift galaxy observations for models.
In particular, we will try to address how likely
such a large concentration of galaxies will occur in 
various models.

We first note that Pettini \etal (1997) give
a selection function (in redshift) of their observations
simply by fitting the observed distribution of galaxies
in the redshift range $z=2-4$.
This approach would be valid, {\it if the distribution of 
galaxies in redshift in the range of interest is constant}.
In general, 
the density of galaxies needs not be constant in redshift.
Without {\it a priori} determined selection function,
we simply use a Gaussian selection function,
$\exp{(z-\bar z)^2/2\sigma_{SF}^2}$,
where the mean $\bar z$ is chosen to be $3.07$
to maximize the selection function value at the peak of 
the observed galaxy concentration ($z=3.05-3.09$)
(this is conservative in the sense that it maximize a model's
ability to match such a large concentration),
and $\sigma_{SF}=0.35$ will be used.
Such a choice of selection function seems
appropriate for the observational technique
used which optimizes detection
at $z\sim 3$ (Pettini 1997, private communications),
and should be adequate 
for our purpose in the sense that our results do not
sensitively depend on the choices of $\bar z$ and $\sigma$
within reasonable ranges.
By the way, this chosen selection window
function turns out to be 
quite close to what is empirically obtained from
observations (Figure 2 of Pettini \etal 1997).

\begin{figure*}
\centering
\begin{picture}(400,260)
\psfig{figure=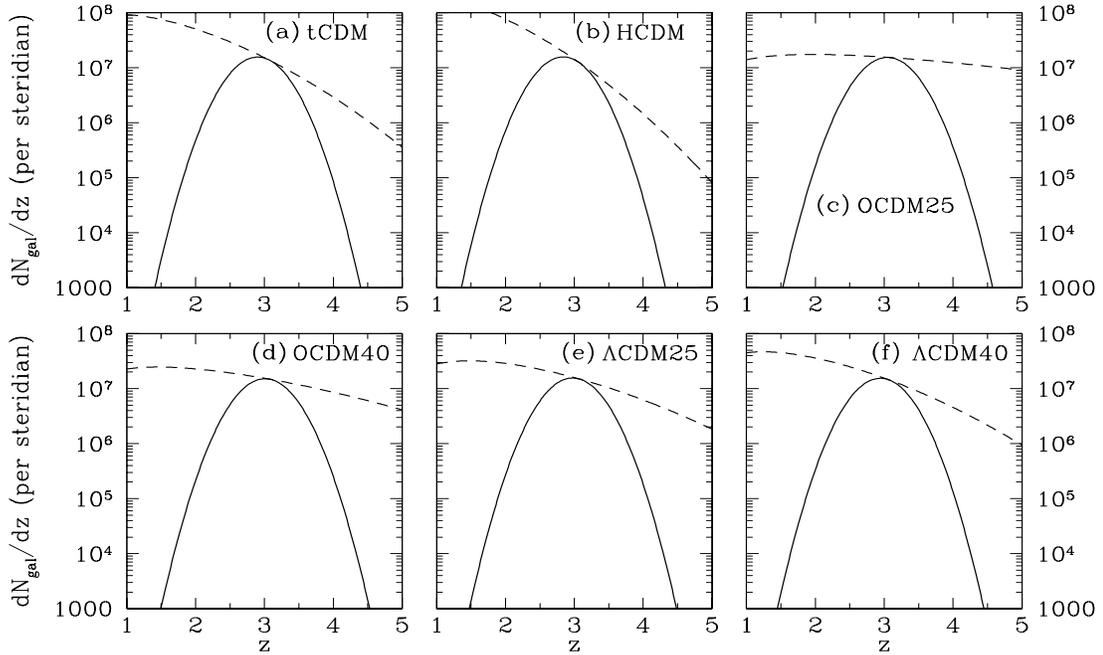,height=10.0cm,width=15.0cm,angle=270.0}
\end{picture}
\caption{
The number of galaxies (halos) per steradian per unit redshift
without (dashed curves)
and with (solid curves) the observational selection function
(see text for the definition of the observational selection function).
The number of galaxies (with the observational selection function)
in each model is normalized to what is observed
in the redshift range $z=2-4$ by Steidel \etal (1997).
}
\end{figure*}

Figure 12 shows the distribution of galaxies (halos)
for each model without (dotted curve)
and with (solid curve) the selection function.
For each model we allow ourselves 
the liberty to adjust the threshold of
the virial mass of the halos until
the number of galaxies in the redshift range $z=2-4$
matches what is observed.
S98 find that $180$ galaxies satisfy
their color selection criteria and about $5\%$
of them turn out to be stars in the field SSA22 of 
size $8'.74\times 17'.64$.
We will simply assume that there are $181\times (1-0.05)=172$ 
observed galaxies in the range $z=2-4$, to which our models
are normalized.
We use GPM (C98) to compute the
abundance of such halos.
Normalizing the number density of galaxies (halos) in each model
requires to impose quite different velocity dispersion threshold 
(the velocity dispersion-mass relation is governed by equation 18
and the virial radius-mass relation by equation 12 of C98)
for the virialized halos as tabulated in Table 3.
Preliminary observational evidence indicates
that the high redshift galaxies 
have velocity dispersions in the range 
$\sigma_{||}=180-320$km/s (Steidel \etal 1996). 
We thus see that the current observed range in $\sigma_{||}$
has already become interesting 
in terms of imposing a constraint on the models.
If future observations can significantly narrow the range,
then the combination of $\sigma_{||}$ and density
may be quite powerful to discriminate between models.
We further note the volume-limited
redshift distributions of galaxies (solid curves
in Figure 12) are quite different in different models.
So a direct observation of this (without narrow observational
selection window functions) in a sufficiently large range in redshift,
perhaps by a combination of several observational techniques/passbands,
will be very important.

\begin{deluxetable}{cccccccccc} %{l,r}
\tablewidth{0pt}
\tablenum{3}
\tablecolumns{7}
\tablecaption{High Redshift Galaxies in Six Models} %\label{tab1}}
\tablehead{
\colhead{Model} &
\colhead{$L_x$ ($h^{-1}$Mpc)} &
\colhead{$L_y$ ($h^{-1}$Mpc)} &
\colhead{$L_z$ ($h^{-1}$Mpc)} &
\colhead{$\sigma_{||}$ (km/s)} &
\colhead{$\sigma_{m,box}$} &
\colhead{$b_{halo}$}}

\startdata
tCDM & $7.63$ & $15.4$ & $15.0$ & $165$ & $0.138$ & $4.82$ \nl 
HCDM & $7.63$ & $15.4$ & $15.0$ & $115$ & $0.131$ & $6.38$ \nl 
OCDM25 & $10.8$ & $21.8$ & $22.7$ & $243$ & $0.320$ & $1.90$ \nl 
OCDM40 & $9.65$ & $19.5$ & $20.2$ & $224$ & $0.239$ & $2.71$ \nl 
$\Lambda$CDM25 & $11.9$ & $24.1$ & $29.3$ & $202$ & $0.207$ & $3.30$ \nl 
$\Lambda$CDM40 & $10.4$ & $20.9$ & $23.4$ & $185$ & $0.178$ & $3.92$ \nl 
\enddata
\end{deluxetable}

\begin{figure*}
\centering
\begin{picture}(400,260)
\psfig{figure=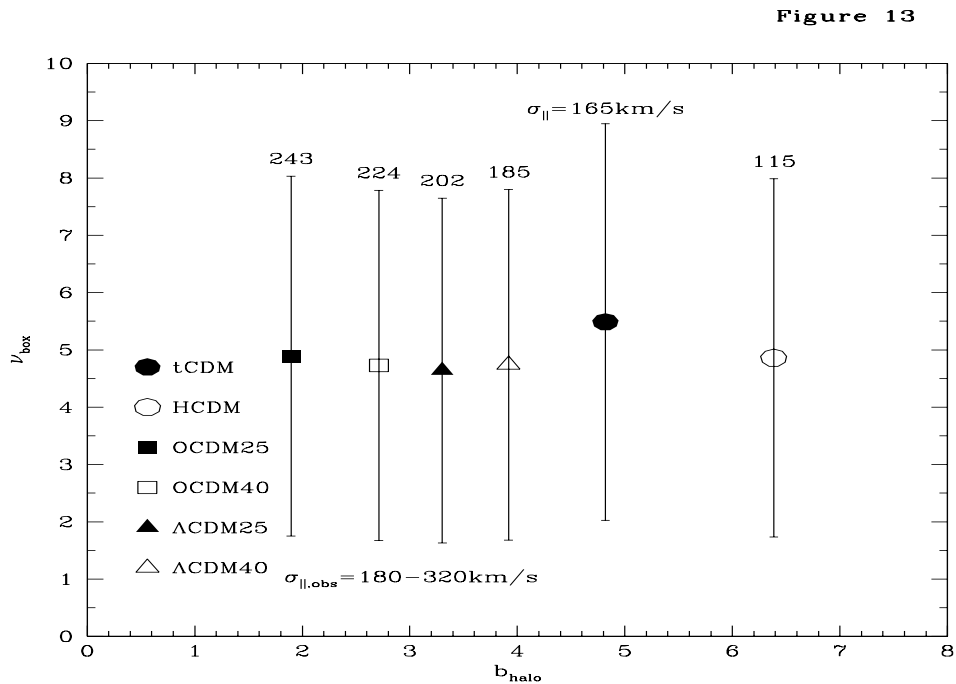,height=10.0cm,width=15.0cm,angle=0.0}
\end{picture}
\caption{
Peak height of the box ($\nu_{halo,box}$) against
the bias of galaxies over the matter on the box ($b_{halo}$)
is plotted
for each model.
The range in vertical axis for each model is ($2\sigma$)
statistical fluctuations due to small number
of galaxies in the observed galaxy peak.
Also shown above each vertical errorbar
is the velocity dispersion threshold of halos in each model,
along with the observed range shown near the bottom of the plot.
}
\end{figure*}

The field of view of size
$8'.74\times 17'.64$ corresponds to different sizes
($L_x$, $L_y$) in comoving length units for different models, 
so does the redshift bin size ($\Delta z=0.04$; $L_z$),
all of which are listed in Table 3 along with
some other relevant quantities.
We compute the rms fluctuations of total mass in a box of 
comoving size $L_x\times L_y\times L_z$ for each model at $z=3$,
$\sigma_{m,box}$,
using the following formula in Fourier space:
\begin{eqnarray}
\sigma_{m,box}^2={1\over (2\pi)^3}\int_0^\infty \int_0^\infty \int_0^\infty \hskip -0.5cm&\left[\sin (0.5 L_x k_x)/(0.5 L_x k_x)\right]^2\nonumber\\
	    &\left[\sin (0.5 L_y k_y)/(0.5 L_y k_y)\right]^2\nonumber\\
	    &\left[\sin (0.5 L_z k_z)/(0.5 L_z k_z)\right]^2 \nonumber\\
            & P_k dk_x dk_y dk_z,
\end{eqnarray}
\noindent 
where $P_k$ is the power spectrum at $z=3$.
Note that we have assumed that 
{\it there is no peculiar velocity 
distortion along the line of sight},
when computing $L_z$.
This assumption is conservative in the sense that
the actual extent in real space would be larger than
the indicated $L_x$, 
if the linear compression along the line of sight for an overdense
region is more important than 
the finger-of-god effect due to the internal velocity dispersion
of the collapsed object.
We have the following reason to believe that this is the case.
Given the observational evidence that such peaks are fairly common
(S98),
it is highly unlikely that they are {\it very rich} clusters of galaxies,
thus internal velocity dispersion is unlikely
to be important on the indicated scales of $L_z$.
To be somewhat more quantitative, a velocity dispersion of
$1000$km/s corresponds to $5h^{-1}$Mpc comoving at $z=3$ in an 
$\Omega_0=1$ model, significantly smaller than $L_z=15h^{-1}$Mpc
in the same model. 

Let us now compute the bias factor of the halos
at $z=3$ using the method similar to that
used to compute the correlation of clusters 
(equation 28 of C98): 
$b_{halo}\equiv {<\tilde\nu>\over \sigma_0} + 1$,
which is listed in Table 3.
$<\tilde\nu>$ is computed using equation 29 of C98
and $\sigma_0$ is the rms fluctuation, both computed
for a density field at $z=3$
smoothed by a Gaussian window chosen to 
produce the correct number density of halos (C98).
We see that peaks are moderately biased over mass in low
density models but strongly biased in $\Omega_0=1$ models,
in agreement with the conclusion drawn by S98.

If a linear bias of halos over mass on the relevant large scales
(i.e., on the box scale) exists,
as supported by detailed 
galaxy formation simulations (Cen \& Ostriker 1992b; 
Blanton \etal 1998) and analytical arguments (Gramann \& Einasto 1992),
one obtains
\begin{equation}
\sigma_{halo,box} = b_{halo}\sigma_{m,box},
\end{equation}
\noindent 
where $\sigma_{halo,box}$ is the rms fluctuations
of halos in a box of comoving size 
$L_x\times L_y\times L_z$ at $z=3$.
We first compute the number of galaxies
(without the selection function)
in each model in the redshift range $z=2.4-3.8$ 
consisting of 35 redshift bins each of size
$\Delta z=0.04$ (Steidel \etal 1998), $N_{2.4-3.8}$.
Dividing $N_{2.4-3.8}$ by $35$
gives the average number of galaxies in each redshift bin of size,
$\bar N$.
Since the total number of spectroscopically identified
galaxies is only $78$ in the redshift range 
$z=2.4-3.8$,
we need to compensate for the fraction of 
spectroscopically yet unidentified galaxies,
which will be assumed here to have
the same redshift distribution as the spectroscopically
identified ones.
This is accounted for by multiplying the relevant
galaxy number by the ratio, $N_{2.4-3.8}/78$.
Now, we can compute 
the height of the observed galaxy peak at $z=3.07$ 
in terms of the rms fluctuations in the indicated 
box ($L_x\times L_y\times L_z$) in each model:
\begin{equation}
\nu_{box}={N_{peak} (N_{2.4-3.8}/78)/\bar N\over \sigma_{halo,box}}-1,
\end{equation}
\noindent 
where $N_{peak}$ is the number of galaxies in the observed
redshift peak near $z=3$.
There are $15$ spectroscopically identified galaxies
in the peak ($\Delta z= 3.05-3.09$) in the SSA22 field,
which gives a $2\sigma$ upper and lower range 
of $N_{peak}=15\pm 8$ due to Poisson fluctuations.

Figure 13 shows 
$\nu_{box}$ as a function $b_{halo}$.
The range in the vertical axis for each model is due to ($2\sigma$)
Poisson statistical fluctuations of the small number 
of galaxies in the observed galaxy peak.
Also shown above each vertical errorbar 
is the velocity dispersion threshold of halos in each model,
along with the observed range shown near the bottom of the plot.
Clearly, the vertical ranges are large.
Hence, a larger sample of such events will be
invaluable to dramatically shrinking the size of the errorbars.
If such galaxy concentrations turn out to
be as common as $2\sigma$ events
in the real world, then all the models would have to be re-evaluated.
The two $\Omega_0=1$ models (tCDM and HCDM) have one additional
inconsistency with observations:
they do not have enough massive halos to host galaxies
with velocity dispersion as large as 
indicated by the preliminary observations.

It should be pointed out that the GPM method used here
to compute the halo statistics ignores merger.
But it may be argued that,
in this case, 
this peak counting method may be more accurate than
direct N-body simulations for precisely the reason
that merger is ignored.
Overmerging of galaxies (identified with halos)
in N-body simulations is error-prone 
due to lack of realistic physics/sufficient resolution.
In real world, although halos may still merge,
most galaxies are likely to retain their separate
identities (for example, a rich cluster like Coma
contains hundreds of galaxies but might consist of only
a handful of dark matter halos).
This is, of course, not in conflict with some
unavoidable processes occurring in the real world
that may transform the morphologies of galaxies but
do not produce direct merger, such as galaxy harassment (Moore \etal 1996).
Thus, our treatment by ignoring merger may in fact
be closer to truth than halos 
identified in direct N-body simulations in terms
of correctly counting galaxies.
If halo merger is more important in objects such as
the observed galaxy concentration in the real world,
inclusion of some merger effects would exacerbate the problem,
as merger would 
reduce the number of galaxies in high density region (as 
considered here) and reduce the bias of galaxies (in number)
over mass in the models.
Interestingly, our results, namely that the high redshift galaxies are
highly biased over mass, 
are in broad agreement with results from N-body simulations
(Bagla 1998; Jing \& Suto 1998; Governato \etal 1998).

We note in passing that
directly measuring the bias of galaxies
over mass at high redshift will prove to be extremely difficult.
The only {\it direct} measurement of the distribution of
underlying mass has to come from
direct mapping by gravitational lensing.
But such a measurement, even if possible in principle,
is very difficult.
This situation is further worsened by the fact
that matter at $z=3$ does not make an optimal
contribution to the lensing effect compared to matter
at lower redshifts.
There is, however, 
one possible, though somewhat less direct,
measurement of the bias of galaxies over mass.
It makes use of the fact
that it is possible to
select a population of $\lya$ clouds whose
clustering strength is comparable to that of the underlying matter.
Then, a comparison between the clustering of such $\lya$ clouds
and the clustering of galaxies both at the same high redshift
would yield a measurement of bias (of galaxies over matter).
This technique has recently been proposed by Cen \etal (1998).
But detailed computations of various models
are needed in order to understand the relationship
between matter and $\lya$ clouds in different models,
before one can make self-consistent calibration
to infer the clustering of mass from that of $\lya$ clouds.

\subsection{Reionization of the Universe}

%We now turn to reionization of the universe.
To properly tackle this issue requires
detailed hydrodynamic modeling with three-dimensional
radiative transfer.
Cosmological hydrodynamic simulations (Cen \& Ostriker 1993a;
Ostriker \& Cen 1996; 
Gnedin \& Ostriker 1997) 
as well as analytic analyses (Haiman \& Loeb 1997) have 
demonstrated that the detailed picture
of reionization of the universe
depends on many uncertain parameters,
including the properties of the ionizing sources,
the clumpiness of the baryonic density field,
the mean baryonic density, etc.

We want to address the following comparatively simpler question:
When was the universe completely reionized?
It is possible to obtain a relatively accurate result
if one only wants to identify the epoch when the universe
gets completely reionized
(ignoring the detailed development of the reionization process).
In this case one just needs to count the number of ionizing photons.
When the number of ionizing photons emitted per hydrogen atom
in one (hydrogen) recombination time is equal to one,
the universe (with regard to hydrogen) is completely ionized.
We now formulate this criterion quantitatively,
assuming that the universe 
is reionized primarily by UV ionizing
radiation from massive stars of early galaxies.

The Hubble time as a function of redshift 
is (Peebles 1993)
\begin{equation}
t_H(z) = t_0\left[\Omega_0(1+z)^3+(1-\Omega_0-\Lambda_0)(1+z)^2 + \Lambda_0\right]^{-1/2},
\end{equation}
\noindent 
where $t_0=9.78h^{-1}~$Gyr is the present Hubble time.
The hydrogen recombination time is
\begin{equation}
t_{rec}=[1.16 n_H \alpha(T) C^2]^{-1} = 2.0 \left({\Omega_b h^2\over 0.0125}\right)^{-1}\left({1+z\over 6}\right)^{-3} C^{-2},
\end{equation}
\noindent 
where $1.16$ comes from the fact that the 
primordial gas has a (76\%, 24\%) (H, He) composition,
$\alpha(T)$ is the hydrogen recombination coefficient
and $C\equiv \sqrt{<\rho_b^2>/<\rho_b>^2}$ is the gas
clumping factor (we have ignored here the difference between 
total baryonic matter and ionized hydrogen assuming that hydrogen
is ionized everywhere).
For the second equality in equation 15 we assume $T=1.5\times 10^4$K,
giving $\alpha=6.9\times 10^{-13}~$cm$^3$/sec (Cen 1992).
Assuming that a fraction, $\epsilon_{UV}$, of the rest mass energy
of collapsed systems is converted into ionizing photons in 
a short time scale (i.e., star formation for the relevant
massive stars is assumed to be almost instantaneous once
a galaxy has collapsed),
we can write down the equation, equating the number of ionizing photons
emitted per hydrogen atom in a recombination time to unity,
as:
\begin{equation}
{\epsilon_{UV} m_p c^2\over h \nu_0}{df_g\over d z}{d z\over d t_H} t_{rec}=1,
\end{equation}
\noindent 
where $df_g/dz$ is the fraction of total baryonic matter that has
collapsed and condensed out to form galaxies per unit redshift,
$m_p$ is proton mass, $c$ is the speed of light
and $h\nu_0$ ($=13.6$eV)
is the hydrogen ionization potential.
Using equations (14,15) to expand equation (16),
one finds
\begin{eqnarray}
{df_g\over d z}&=&1.2\times 10^{-9} h^{-1}\epsilon_{UV}^{-1} \left({\Omega_b h^2\over 0.0125}\right)\left({1+z\over 6}\right)^{1/2} C^{2}\nonumber\\ 
&&\left[\Omega_0+(1-\Omega_0-\Lambda_0)(1+z)^{-1} + \Lambda_0(1+z)^{-3}\right]^{-3/2}\nonumber\\ 
&&\left[\Omega_0+{2\over 3}(1-\Omega_0-\Lambda_0)(1+z)^{-1}\right].
\end{eqnarray}
\noindent 
The main uncertainties in this treatment include:
the star formation history of the high redshift galaxy,
the initial mass function (IMF) of the stars,
and the fraction of UV ionizing photons
that escape host galaxies and enter the inter-galactic 
medium (IGM),
the clumpiness of the gas 
and complicated interplay between radiation self-shielding
of dense, optically thick regions and absorption of 
UV photons by these regions.
An accurate treatment of this problem
awaits detailed galaxy formation simulations
with a sufficiently realistic treatment of three-dimensional 
radiative transfer.
The present study serves to illustrate the large differences
between models with regard to the reionization epoch, which
may survive detailed, more accurate simulations.
In the present treatment all 
these various factors have been conveniently
condensed into the single parameter $e_{UV}$,
being the fraction of the rest mass of stars that 
turn into ionizing UV photons escaping into the IGM. 
We will use $\epsilon_{UV}=(3\pm 2)\times 10^{-5}$,
which is lower than the values previously used 
(Gnedin \& Ostriker 1997; Cen \& Ostriker 1998)
by approximately a factor of two.
This factor of two is intended to account for the slight
over-counting of the number of ionizing photons in equation 16,
where all ionizing photons are treated to be at
Lyman limit, in contrast to an Orion-like
spectrum of a young stellar association (Scalo 1986).
A clumping factor $C=7.0$ is used, which may be
uncertain by a factor of a few,
perhaps in the upward direction.

\begin{figure*}
\centering
\begin{picture}(400,300)
\psfig{figure=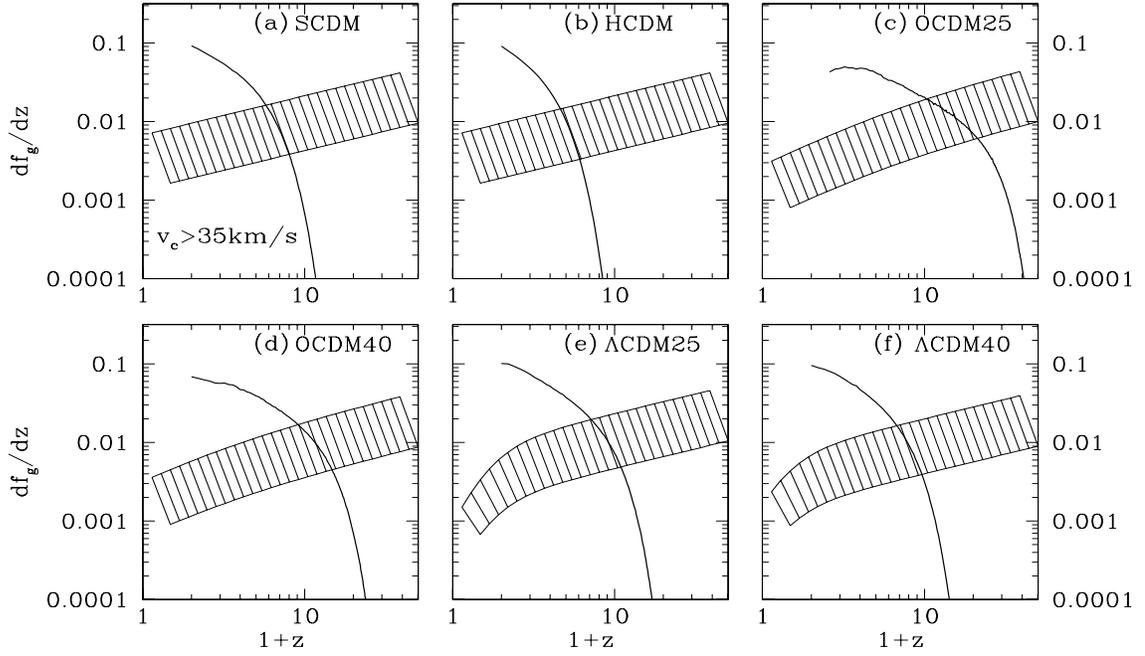,height=10.0cm,width=15.0cm,angle=270.0}
\end{picture}
\caption{
The fraction of matter formed per unit redshift
in collapsed (virialized) systems
with velocity dispersion greater than
$35$km/s (corresponding to virial temperature of $\sim 10^4$Kelvin),
as a function of redshift in six models.
Each cross-shaded band indicates the required
range with which enough UV photons are produced to 
completely reionize the entire universe.
}
\end{figure*}

The required galaxy formation rate per unit redshift
$df_{g}/dz$ (equation 17) is shown as 
the cross-shaded bands in Figure 14,
where the upper and lower limits are due to the uncertainty
in $\epsilon_{UV}$. 
The solid curves in Figure 14 show the 
fraction of matter in collapsed (virialized) systems per unit redshift
with velocity dispersion greater than
$35$km/s (corresponding to virial temperature of $\sim 10^4$Kelvin),
as a function of redshift,
that are assumed to emit the ionizing photons.
Smaller systems are unlikely to be able to collapse
due to the lack of efficient coolants
at $T<10^4$K in the presence of significant 
photo-ionizing radiation field (Efstathiou 1992;
Quinn, Katz, \& Efstathiou 1996) or in the presence of
feedback processes from young galaxies (Cen \& Ostriker 1993a);
molecular hydrogen could have served
as an efficient coolant at the lower temperatures but
they are likely to have been destroyed 
long before these sub-galactic size objects start to form
(Ostriker \& Gnedin 1996; Haiman, Rees, \& Loeb 1997).

\begin{deluxetable}{cccccccccc} %{l,r}
\tablewidth{0pt}
\tablenum{4}
\tablecolumns{2}
\tablecaption{Reionizaton Epochs in Six Models} %\label{tab1}}
\tablehead{
\colhead{Model} &
\colhead{$z_{rei}$}}

\startdata
tCDM & $5.8-8.2$  \nl 
HCDM & $3.7-5.6$  \nl 
OCDM25 & $28.0-36.0$\nl 
OCDM40 & $13.0-20.0$\nl 
$\Lambda$CDM25 & $10.0-13.0$\nl 
$\Lambda$CDM40 & $7.7-10.0$ \nl 
\enddata
\end{deluxetable}

\begin{figure*}
\centering
\begin{picture}(400,270)
\psfig{figure=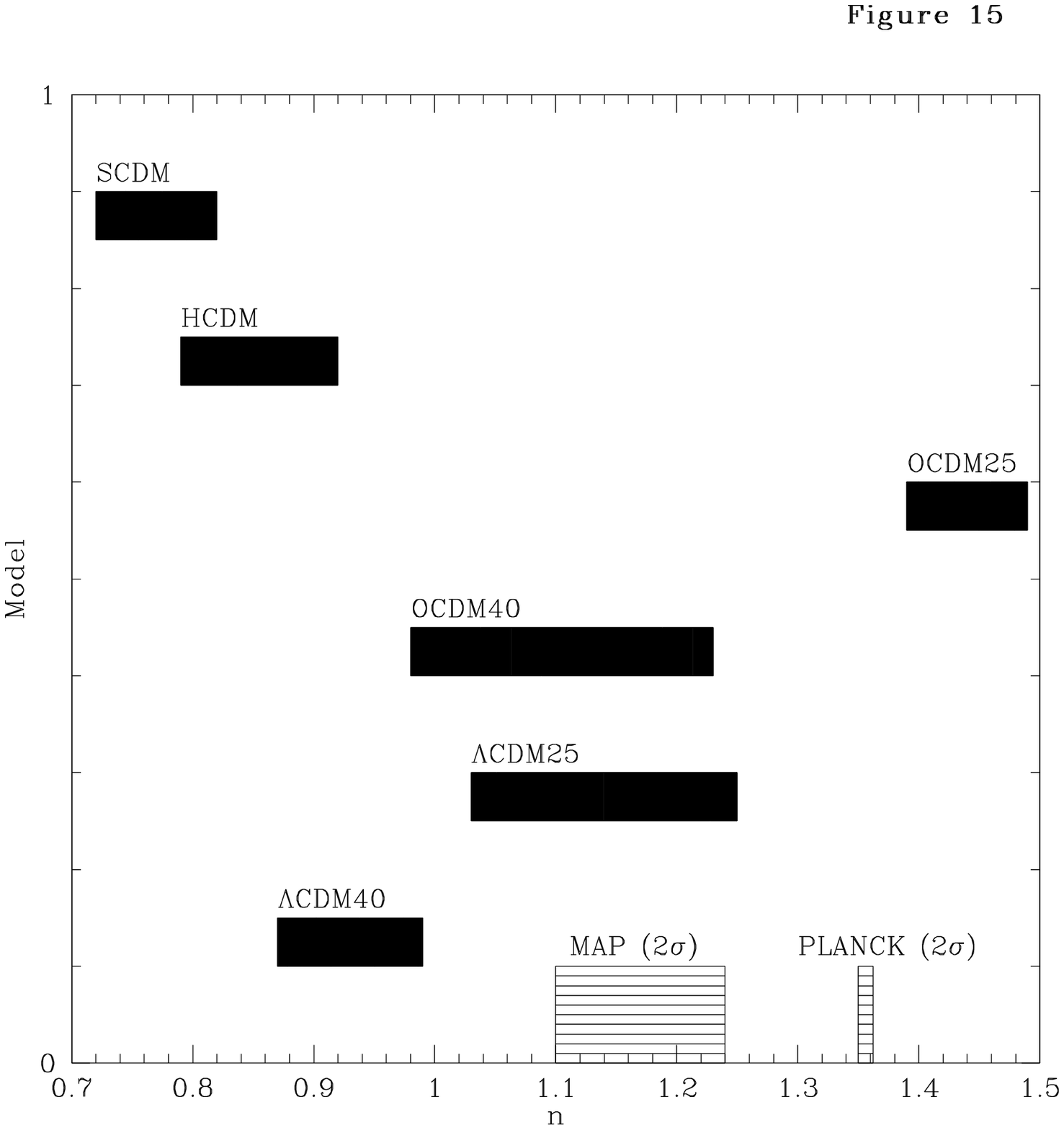,height=10.0cm,width=15.0cm,angle=0.0}
\end{picture}
\caption{
summarizes the model results on $n$ when
all the models are normalized to both COBE and local clusters,
assuming that the primordial spectrum has a pure power law shape
for the scales of cosmological interest ($\sim 1-6000h^{-1}$Mpc).
The ranges in $n$ are $2\sigma$.
Also shown are the $2\sigma$ errorbars that MAP
and PLANCK are thought to be able to achieve
(Zaldarriaga, Spergel, \& Seljak 1997).
Note that the horizontal locations
of MAP and PLANCK indicators are arbitrary.
}
\end{figure*}
We see that different models predict
quite different epochs of reionization of the universe,
which are tabulated in Table 4 
for the ease of comparison.
It is noted that the exact epoch of reionization
is quite insensitive to small variations 
of $e_{UV}$ or the degree of clumpiness of the medium,
simply because it is during the rapidly rising part of the evolution 
of the collapse fraction 
when the reionization occurs.
It is clear that observations to higher redshift
would provide a powerful discriminator between models.
For example, if the universe is observed to be
transparent at $z>5$, then HCDM
will be inviable;
if the universe is observed to be transparent at $z>10$,
the all but the open models will be ruled out.

While it is beyond the scope 
of current paper to conduct detailed calculations of
the properties of high redshift objects around the reionization epoch
in each model,
we only briefly comment that near future observations
by meter waves telescopes such as 
Giant Meterwave Radio Telescope (GMRT; Swarup 1984)
and the Kilometer Square Array Interferometer (SKAI; Braun 1995)
and the next generation infrared telescopes such as 
the Space Infrared Telescope Facility (SIRTF; Werner \& Eisenhardt 1988)
and Next Generation Space Telescope (NGST)
will shed light on the properties of
objects at the reionization epoch such as
neutral hydrogen condensates 
(Scott \& Rees 1990; Subramanian \& Padmanabhan 1993; 
Madau, Meiksin, \& Rees 1997)
and ionizing sources (Haiman \& Loeb 1997; Miralda-Escud\'e \& Rees 1997).
In addition, the differences in the reionization time thus
in optical depth to CMB photon and in angular scales 
would leave different imprints on CMB at small 
angular scales.

\subsection{CMB Experiments: MAP and PLANCK}

Needless to say,
future CMB experiments are likely to 
provide a very rich amount of information, that
can be used in many aspects to constrain cosmological models
as well as understanding astrophysical processes involved.
However, cosmological parameters determined 
from CMB experiments alone are likely to display 
certain degeneracies (e.g., Efstathiou \& Bond 1998).
Other astronomical and cosmological phenomena discussed 
in the preceding sections are, in this sense,
complementary to CMB observations.
Here we choose to focus on one 
parameter, which is cosmologically important but is not
tangled with other complicated processes such as recombination
and reionization
processes, namely, the primordial power index $n$.
Another reason for favoring the choice of $n$
is that $n$ is distinctly different between models 
and perhaps
offers one of cleanest tests of models.
The underlying reason that different models have distinctly different
$n$ is a product of combined effort:
combining CMB (COBE) with local cluster observations.
It is clearly seen from Table 1 that
normalizing a model both to very large scale (COBE)
and cluster scale results in
a tight constraint on the 
primordial power spectrum index $n$.
However, currently COBE does not provide strong constraint on $n$
due to its large errorbars [$\Delta n=0.6$ ($2\sigma$)].
In the foreseeable future 
there are two satellite missions, MAP and PLANCK,
which are proposed to produce much higher resolution
(arminute to sub-degree versus $\sim 10^o$ of COBE) and much more sensitive CMB 
temperature fluctuation measurements.

We summarize the model results on $n$ in Figure 15.
The ranges in $n$ are $2\sigma$.
Also shown are the $2\sigma$ errorbars that MAP
and PLANCK are thought to be able to achieve
(Zaldarriaga, Spergel, \& Seljak 1997).
Note that the horizontal locations 
of MAP and PLANCK indicators are arbitrary.
It is seen that MAP will be able to provide 
a powerful discriminator between the models,
whereas
PLANCK should be able to constrain $n$
thus test models with exquisite precision.

\section{Discussion}

When models are normalized to both COBE on very large scales
and clusters of galaxies on intermediate-to-large scales,
They naturally have 
grossly similar properties of clusters of galaxies at 
zero redshift.
Other local tests surrounding clusters regions or weighted heavily
by clusters are unlikely to be very sensitive for 
the purpose of differentiating between models.
Among such tests are
galaxy-galaxy pairwise velocity dispersion 
(e.g., \cite{dp83}; Strauss, Ostriker, \& Cen 1998),
galaxy-galaxy two point correlation function 
(\cite{dg76}; \cite{p80})
and large-scale velocity flows in the local universe
(e.g., Strauss \etal 1995).
Although it is conceivable that fine details within the clusters at
zero redshift may differ among different models, such as cluster morphology,
substructure etc. 
(Richstone, Loeb, \& Turner 1992; \cite{mefg95};
see Cen 1997b for a somewhat more realistic view
on this considering some unavoidable projection effects),
it is likely that more significant differences between the models
will come from regions far removed from local clusters of galaxies.
There are two ways to be distant from clusters.
First, one can move to regions such filaments or voids
in the local universe
which are spatially distant from the clusters.
There is no doubt that there are significant differences in 
the properties of these regions among different models.
But detailed modeling of such regions, in particular, voids,
is extremely difficult due to the lack of appropriate 
small scale resolution
in such regions even with the best simulations available.
We have followed on the second route in this paper
by traveling back 
in time away from the local universe
and studied the properties of model universes
at moderate-to-high redshift through a suite of cosmological phenomena.

In addition to a consideration of galaxy power
spectrum from SDSS galaxy redshift survey (\S 3.2),
which is shown to be capable of differentiating between models,
we have examined eight phenomena
ranging in redshift from zero to that of last scattering.
They include 
the correlation function of very rich clusters of galaxies (\S 3.1),
the evolution of the abundance of rich clusters of galaxies (\S 3.3),
the gravitational lensing by moderate redshift clusters of galaxies (\S 3.4), 
the properties of $\lya$ forest at high redshift (\S 3.5),
the damped $\lya$ systems at high redshift (\S 3.6),
the high redshift galaxies (\S 3.7),
the reionization epoch of the universe (\S 3.8)
and 
the future high resolution, high sensitivity CMB experiments (\S 3.9).
Some of these observations have already provided
discriminating tests between some models.
For example, HCDM model is strongly ruled out by the observed
large density of neutral hydrogens in damped $\lya$ systems 
and possibly inconsistent with observations of $\lya$ clouds.
Both $\Omega_0=1$ tCDM and HCDM models are
in disagreement with 
the observed abundance of rich clusters of galaxies at $z\sim 0.4-0.8$
and 
the observed strong clustering of high redshift galaxies
at $z\sim 3$, while other, low density models are in fair
agreement with those observations.

We find that all these moderate-to-high redshift 
phenomena possess potentially very powerful 
capabilities of differentiating between all the remaining
``viable" models.
We now discuss the requisite condition for each phenomenon
to be capable of providing some useful constraint.

It is found that different models differ most for correlations
of very rich clusters of galaxies, those with
mean separation $\ge 90h^{-1}$Mpc.
Let us take the data point shown as a solid dot in Figure 3
as an example to quantify this.
If the errorbar were $2h^{-1}$Mpc ($1\sigma$)
(with the mean value unchanged),
all the models may be ruled out at $4\sigma$ confidence level.
Is this achievable? The answer is a probable yes with SDSS.
Let us assume that the correlation function is
a powerlaw with a slope of $-1.8$.
Then, the mean correlation function has
a value $\bar\xi(r_0)=3\xi (r_0)/(3-1.8)=2.5$ 
at $r_0$ (where the correlation function has a value unity).
In order to achieve a statistical errorbar of $2h^{-1}$Mpc 
for $r_0=42h^{-1}$Mpc,
$N_{pair}=(42/2)^2=442$ pairs of clusters are necessary.
Since $\bar\xi=N_{pair}/(V_{survey} n) - 1= 2.5$ at $r_0$,
we get $V_{survey}=N_{pair}/3.5 n = 1.0\times 10^8h^{-3}$Mpc$^3$
[$n=(1/94)^3h^3$Mpc$^{-3}$ is assumed].
The SDSS survey volume is about
${4\pi\over 3} 500^3=5.2\times 10^8h^{-3}$Mpc$^3$.
So it seems that clusters from SDSS will
provide an unprecedented accuracy for the determination
of cluster correlations of very rich clusters.

The evolution rates of the abundances of rich clusters of galaxies
in different models are quite different and 
differences of the abundances of clusters increasingly
deviate from one another towards high redshift (Figure 7).
As shown earlier, even the currently available very small data 
set is already able to differentiate the $\Omega_0$ models (tCDM,HCDM)
from the rest. 
However, larger cluster samples will be required to 
constrain $\Omega_0$ to a high accuracy.
As shown \S 3.3,
a high redshift X-ray cluster survey covering a quarter of the sky
with enough sensitivity to measure
the temperature of hot luminous clusters ($kT\ge 7$keV) at $z\ge 1$
will be
able to determine $\Omega_0$ with an uncertainty
of $\Delta\Omega_0=0.02$ ($1\sigma$), if $\Omega_0+\Lambda_0=1$,
and $\Delta\Omega_0=0.008$, if $\Lambda_0=0$.
These accuracies are comparable to those
which will be determined from the next generation
microwave background fluctuation experiments 
(e.g., Zaldarriaga, Spergel, \& Seljak 1997),
thus provides a very powerful cosmological test and also
a very important consistency cross-check.

The gravitational lensing optical depths 
for large splittings or giant arcs 
due to moderate-to-high redshift clusters of galaxies
increasingly differ towards high redshift between models. 
While this is closely related to the evolution of abundance of rich
clusters, the differences between different models tend
to be larger and more extreme because 
of the highly nonlinear
dependence of gravitational lensing cross section on cluster mass.
For background sources at $z\le 1$,
the primary differences occur between models
with different $\Omega_0$;
models with a same $\Omega_0$ but with and without
a cosmological constant 
give comparable results (Figure 9a).
For background sources at $z\sim 3$,
all the models become non-degenerate (Figure 9b).
A systematic search of about $10^6$ background sources at $z\ge 3$
will be desirable.
In fact, SDSS should provide such a sample ($\sim 10^6$ quasars mostly
at high redshift with photometric redshifts and $10^5$ quasars with
spectroscopic redshifts, Margon 1998).
One can use such a sample in two ways.
Firstly,
the total number of lensed sources will be different between the models.
With $10^6$ quasars 
(assuming to be at $z_s=3$ for this simple illustration),
one should see 
($1900\pm 44$, $710\pm 27$, $570\pm 24$, $330\pm 18$, $100\pm 10$, $39\pm 6$)
multiple-image quasars with separation $\ge 30^"$
in (OCDM25, OCDM40, $\Lambda$CDM25, $\Lambda$CDM40, tCDM, HCDM),
respectively.
This would allow an unambiguous test of these six models.
It would put a constraint on 
$\Omega_0$ with $\Delta\Omega_0\sim 0.01$ if the
universe is open or 
$\Delta\Omega_0\sim 0.03$ if the universe is flat.
Secondly, and more sensitively,
the distributions of lenses themselves 
in different models will be very different.
For example, one expect to see 
(200.0, 4.0, 0.5, 0.005, 0.0, 0.0) lenses
at $z>2$ in 
(OCDM25, OCDM40, $\Lambda$CDM25, $\Lambda$CDM40, tCDM, HCDM),
respectively, that produce arcs or splittings $\ge 30^"$.

The properties of $\lya$ forest at high redshift are
likely to be dramatically different among the models given
the order of magnitude different powers on the relevant scales.
With regard to this, observations are perhaps slightly ahead of theory;
there is a large and still rapidly accumulating observational database
for $\lya$ forest.
But detailed hydrodynamic simulations of $\lya$
forest, while having made ground-breaking advances in our understanding
of $\lya$ forest phenomenon in recent years,
are yet to sort out the fine dependence
of results on such issues as variations among different hydrocodes,
resolution and boxsize effect, radiative transfer effect
and star formation feedback effects,
before they can be utilized to test cosmological models
in accurate (thus meaningful) ways.
Preliminary comparisons between models and observations
using the one-point flux distribution probability function
(Rauch \etal 1997)
favor the two $\Lambda$ based models 
($\Lambda$CDM25 and $\Lambda$CDM40),
but disfavor HCDM and OCDM25 in the opposite sense that
HCDM has too little power and OCDM25 has too much power
on the relevant scales.
tCDM and OCDM40 appear to be marginally consistent.

The neutral hydrogen mass in damped $\lya$ systems at high redshift is
undoubtedly a very sensitive test.
Their role is comparable to that of rich clusters today
in the sense that 
damped $\lya$ systems are rare systems at the high redshift.
Four improvements will dramatically enhance its discriminatory power.
First, more detailed kinematic studies of these systems,
at the highest redshift possible,
are indispensable to providing more tight
limits on the mass of these systems.
Second, a significant reduction of observational errorbars with
more samples can significantly increase the sensitivity of the test.
SDSS should provide a much larger quasar sample.
Third, it is beneficial to go to higher redshift to further
increase its sensitivity of testing cosmological models.
Finally, detailed modeling of these systems 
is necessary in order to determine the precise
amount of gas in neutral form.
For this we will have to implement detailed three-dimensional
radiative transfer in the existing hydrocodes,
run large simulation boxes ($\ge 20h^{-1}$Mpc) 
to ensure a fair sample
and have a better understanding of the star formation processes
in these systems.

The abundance and
clustering of high redshift galaxies at $z\sim 3$
has been shown to differ among models.
These are the large proto-clusters or proto-superclusters
that eventually produce the clusters and superclusters of galaxies
that we see today.
The issue here is to enlarge the observational sample size.
If the observed galaxy concentration seen by 
S98 turns out to be ubiquitous,
all the models considered would be in a very
uncomfortable position.
With about ten fields like SSA22, preferably
randomly distributed spatially,
we may be able to reject some models with high confidence levels.
A perhaps more discriminatory test with the high redshift galaxies
is from measuring the velocity dispersion (or mass) 
of these galaxies.
If the high redshift galaxies are as massive 
as the present day $L^*$ galaxies,
as present preliminary observational evidence indicates (Steidel \etal 1996),
all $\Omega_0=1$ models are inviable.

The reionization epoch of the universal baryons 
is shown to be substantially different among the models.
It is very worthwhile to search for quasars
at as high redshift as possible.
For example, if quasars can be seen beyond
$z>5$, HCDM will be inviable;
if quasars can be seen beyond
$z>10$, only low density open models could pass.
Next generation infrared telescopes such as SIRTF and NGST
and next generation radio telescopes such as GMRT and SKAI
will prove to be extremely useful to not only 
test the models but also understand the detailed history
of the reionization. 
Future CMB experiments will also shed light on this.

Finally,
the future high resolution, high sensitivity CMB experiments
such as MAP and PLANCK will provide many sensitive tests.
Among them one critical test will be an unambiguous determination
of the primordial power index.
Accurate determinations by MAP and PLANCK will provide strong 
constraints on $\Omega_0$.

\section{Conclusions}

We examine a set of
observations and compare them (if currently available)
with a set of six COBE and cluster normalized CDM 
models, or forecast their capabilities in constraining models
or outline the requirements for them to be capable of constraining models
in a dramatic fashion.
Detailed discussion has been presented in the previous section.
In addition to SDSS galaxy redshift survey (\S 3.2),
we have examined eight different phenomena
including
the correlation function of very rich clusters of galaxies (\S 3.1),
the evolution of abundance of rich clusters of galaxies (\S 3.3),
the gravitational lensing by moderate-to-high redshift clusters of galaxies
(\S 3.4),
the properties of $\lya$ forest (\S 3.4),
the damped $\lya$ systems at high redshift (\S 3.6),
the high redshift galaxies (\S 3.7),
the reionization epoch of the universal baryons (\S 3.8)
and
the future high resolution, high sensitivity CMB experiments (\S 3.9).
Some of these observations have already provided
discriminating tests on some models.

The hot+cold dark matter model (with $\Omega_h=0.2$)
is already ruled out at a $>3\sigma$ level
by current observations of damped $\lya$ systems at high redshift
and the observed high abundance of rich clusters of galaxies at $z\sim 0.4-0.8$.
There is some preliminary indication that there is not enough power
in HCDM to account for the observed high redshift galaxies
and the $\lya$ forest lines.
HCDM is marginally consistent with the existence
of $z\sim 5$ quasars in the sense that the HCDM universe
would be marginally completely reionized by $z\sim 5$;
seeing higher redshift quasars will be an embarrassment to HCDM.
Overall, HCDM is the least viable model among examined
and is ruled out in several independent ways at high confidence levels.

The tilted CDM model (tCDM) is 
in disagreement with 
the observed high abundance of rich clusters of galaxies at $z\sim 0.4-0.8$
at a high confidence level ($>3\sigma$).
It is inconsistent with the observed
strong correlation of clusters of 
galaxies --- another manifestation of lack of large-scale
power of the model 
(even with a substantial tilt, as is required to
fit both COBE and local cluster abundance and included in the analysis).
tCDM is marginally inconsistent ($\sim 1-2\sigma$) 
with the observations 
of damped $\lya$ systems in the sense that it underpredicts
the amount of neutral hydrogen in damped $\lya$ systems 
at $z\sim 3-4$,
but is marginally consistent with the observed flux distribution
of $\lya$ forest.
In summary, tCDM does not seem to be favored.

Low density, flat CDM models with a cosmological constant 
are consistent with all the observations examined here
with the preferred range of the density parameter being
$\Omega_0=0.30\pm 0.10$. 
This preferred range is primarily constrained at the high end 
by the abundance of clusters of galaxies at $z\sim 0.5$ 
(in order not to underproduce the cluster abundance)
and at the lower end by the flux distribution of $\lya$ forest at $z\sim 2$
(in order not to overproduce the fluctuations in the flux distribution)
and also by observations of the damped $\lya$ systems $z\sim 3$
(in order not to overproduce the abundance of damped $\lya$ systems).

Low density, open CDM models 
are consistent with all the observations examined that
are currently available with
the preferred density range being 
$\Omega_0=0.45\pm 0.10$,
which is constrained
by similar observations aforementioned in the preceding 
paragraph for the $\Lambda$CDM models.
An additional constraint is provided 
by the observed galaxy fluctuations in the local universe,
$\sigma_8(gal)\sim 1.0$.
Since it is improbable for galaxies to be anti-biased 
tracers of the mass, $\sigma_8(gal)\sim 1.0$ (Strauss \& Willick 1995) 
implies $\sigma_8(mass) \le 1.0$, which 
translates into an requirement $\Omega_0\ge 0.25$ for open CDM models.

Some improvements on the comparisons between models
and observations will be likely in a relatively short term,
either through more realistic modeling and/or by improved observational
data, including the correlation function of very rich clusters
of galaxies at z=0 (from SDSS),
the $\lya$ forest at $z\sim 2-4$ (primarily by more careful
modeling and analysis),
damped $\lya$ systems (by Keck) and high redshift galaxies at $z\sim 3$
(using Keck and others).
These observations look at different physical scales of a model
at different redshift:
the 
correlation function of very rich clusters of galaxies
probes the power on large scales,
$\lambda \ge 100h^{-1}$Mpc, at $z=0$;
the $\lya$ forest lines measure the power on small scales,
$\lambda \sim 0.1h^{-1}$Mpc, at $z\sim 2-4$;
the damped $\lya$ systems check the power on intermediate scales,
$\lambda \sim 1.0h^{-1}$Mpc, at $z\sim 3$;
and 
the abundance and clusters of 
the high redshift galaxies constrain the power on intermediate-to-large
scales,
$\lambda \sim 1.0-100h^{-1}$Mpc, at $z\sim 3$.
These improvements will further tighten the parameter space
for the low density models.

The evolution rates of the abundances of rich clusters of galaxies
in different models are quite different and 
differences of the abundances of clusters increasingly
deviate from one another towards high redshift.
A large and deep X-ray cluster survey would 
be invaluable to determine the abundance of high redshift,
hot and luminous X-ray clusters, which would permit an 
accurate determination of $\Omega_0$.
A large survey covering about a quarter of the sky
and capable of detecting hot and luminous clusters ($kT\ge 7$keV)
at $z\ge 1.0$
would place a constraint on $\Omega_0$ with $\Delta\Omega_0\sim 0.01$.
These accuracies are comparable to those
which will be determined from the next generation
microwave background fluctuation experiments 
(e.g., Zaldarriaga, Spergel, \& Seljak 1997),
thus provide a very powerful test and also
a very important consistency cross-check.
While AXAF will provide a deep sample of X-ray clusters of galaxies,
a larger survey is required to make a competitively
accurate constraint on $\Omega_0$.

Gravitational lensing observations of giant arcs or very large
splitting multiple quasars by clusters of galaxies
will provide very sensitive tests.
For the same reason that cluster abundances at high redshift
are different in different models, gravitational lensing
provides an independent measurement of the same objects -- rich 
clusters of galaxies at moderate to high
redshift -- as well as lensed high redshift sources.
The quasar data set from SDSS may permit an unprecedented,
accurate determination of both the lensing optical depth  
and the redshift distribution of the lenses themselves
(perhaps in conjunction with follow-up observations to determine
the redshifts of high redshift lenses).
A very tight constraint can be placed on $\Omega_0$ with
$\Delta\Omega_0\sim 0.01$.

Observation of the reionization of the universe would 
provide the potential to differentiate between open and $\Lambda$
CDM models. 
Perhaps the simplest discriminator will be to sight
a very high redshift quasar $z>10$.
If such a quasar is seen, open models will be preferred.

A combination of several (or all) of the aforementioned observations
will offer several ways to 
break the current degeneracy of models that are both
COBE and cluster normalized,
primarily between open CDM and $\Lambda$ CDM models.
On the other hand,
any conflict among the observations with respect 
to all the models may have profound 
implications 
with respect to the Gaussian picture of structure formation 
and/or the general gravitational instability theory.

\acknowledgments
The work is supported in part
by grants NAG5-2759 and ASC93-18185.
Discussions with 
Drs. \hskip -0.2cm N. Bahcall, J.R. Bond, M. Dickinson, G. Lake, J.P. Ostriker,
D.N. Spergel and M.A. Strauss are gratefully acknowledged.
D. Goldberg and M. Strauss are thanked for supplying the 
SDSS power spectrum results.
Special thanks are extended to an anonymous
referee who made several helpful comments.
Finally, I would like to thank 
George Lake and University of Washington for
the warm hospitality, and financial support
from the NASA HPCC/ESS Program during a visit when
this work was initiated.

\end{document}